\documentclass[11pt,a4paper]{article}

\usepackage{bm,bbm,amsfonts,amssymb,amsmath,color}
\usepackage[mathscr]{euscript}
\usepackage{graphicx}
\usepackage{hyperref}
\usepackage{tikz}

\newcommand{\bC}{\mathbb{C}}

\newcommand{\bR}{\mathbb{R}}
\newcommand{\bZ}{\mathbb{Z}}
\newcommand{\bH}{\mathbf{H}}
\newcommand{\bK}{\mathbf{K}}
\newcommand{\bN}{\mathbf{N}}
\newcommand{\bM}{\mathbf{M}}
\newcommand{\bQ}{\mathbf{Q}}
\newcommand{\bV}{\mathbf{V}}
\newcommand{\bX}{\mathbf{X}}
\newcommand{\bY}{\mathbf{Y}}

\newcommand{\cN}{\mathcal{N}}
\newcommand{\cO}{\mathcal{O}}

\newcommand{\cX}{\mathcal{X}}

\newcommand{\sA}{\mathsf{A}}
\newcommand{\sP}{\mathsf{P}}

\newcommand{\sT}{\mathsf{T}}

\newcommand{\si}{\mathsf{i}}

\newcommand{\sY}{\mathsf{Y}}

\newcommand{\frkg}{\mathfrak{g}}

\newcommand{\frkq}{\mathfrak{q}}
\newcommand{\msA}{\mathscr{A}}
\newcommand{\msF}{\mathscr{F}}
\newcommand{\msS}{\mathscr{S}}
\newcommand{\msY}{\mathscr{Y}}
\newcommand{\msW}{\mathscr{W}}
\newcommand{\Tr}{\operatorname{Tr}}

\newcommand{\Li}{\operatorname{Li}}

\newcommand{\ind}{\mathbb{I}}
\newcommand{\rn}{{\mathrm{n}}}
\newcommand{\be}{\begin{equation}}
\newcommand{\ee}{\end{equation}}
\newcommand{\tmu}{\tilde{\mu}}
\newcommand{\tx}{\tilde{x}}
\newcommand{\bea}{\begin{eqnarray}}
\newcommand{\eea}{\end{eqnarray}}
\newcommand{\nn}{\nonumber}

\newcommand{\trn}{\tilde{\rm{n}}}

\newcommand{\ci}{\iota}

\newcommand{\bra}[1]{\langle #1 |}
\newcommand{\ket}[1]{| #1 \rangle}
\newcommand{\vev}[1]{\langle #1 \rangle}

\makeatletter

\@addtoreset{equation}{section}
\makeatother

\begin{document}

\thispagestyle{empty}
\begin{flushright} \today \end{flushright} 

\vskip2cm
\begin{center}
 \hspace{-1cm}%
 {\Large \textbf{Quantum integrability from non-simply laced quiver gauge theory}}
 \vskip1.5cm
{\large 
 \textsc{Heng-Yu Chen$^1$ and Taro Kimura$^2$}}

 \vskip1.5cm

 $^1$\textit{Department of Physics and Center for Theoretical Sciences,\\
 National Taiwan University, Taipei 10617, Taiwan}

 \vskip.5cm
 $^{2}$\textit{Department of Physics, Keio University, Kanagawa 223-8521, Japan}
\end{center}

\vskip1cm
 \begin{abstract} 
  We consider the compactifcation of 5d non-simply laced fractional quiver gauge theory constructed in~\cite{Kimura:2017hez}.
  In contrast to the simply laced quivers, here two $\Omega$-background parameters play different roles, so that we can take two possible Nekrasov--Shatashvili limits.
  We demonstrate how different quantum integrable systems can emerge from these two limits, using $BC_2$-quiver as the simplest illustrative example for our general results.
  We also comment possible connections with compactified 3d non-simply laced quiver gauge theory.
 \end{abstract}

\renewcommand{\thefootnote}{\arabic{footnote}}
\setcounter{footnote}{0}

\vfill\eject

\tableofcontents

\hrulefill

\vspace{1em}

\section{Introduction and Conclusion}
\paragraph{}
The moduli spaces of vacua in 4d $\cN=2$ gauge theory exhibit a fascinating correspondence to the algebraic classical integrable systems~\cite{Gorsky:1995zq,Martinec:1995by,Donagi:1995cf,Seiberg:1996nz}.
The Coulomb branch of the moduli space for the compactified gauge theory can be identified with the phase space of the associated classical integrable system. 
Under this correspondence, the chiral ring operators in gauge theory are interpreted as the commuting conserved Hamiltonians on the integrable system side, and thus the Seiberg--Witten curve in gauge theory is then identified with the spectral curve of the classical integrable system that generate such operators.
\paragraph{}
This correspondence is nicely lifted to the quantum correspondence:
It was shown by Nekrasov--Shatashvili (NS)~\cite{Nekrasov:2009rc} that equivariant deformation of gauge theory, in particular, by turning on one of the equivariant parameters $(\epsilon_1,\epsilon_2)$ as $\epsilon_1 = $ fixed, $\epsilon_2 \to 0$, gives rise to the correspondence to quantum integrable systems.
In this correspondence, the twisted $F$-term condition is identical to the Bethe equation of the associated quantum integrable system, where the quantization parameter is given by $\hbar = \epsilon_1$, and the quantum analog of the Seiberg--Witten curve, which is a difference equation instead of an algebraic relation, is translated to the TQ-relation of the quantum integrable system.
\paragraph{}
For example, under this identification, 4d $\cN=2$ $U(n)$ gauge theory with $n^\text{f} = 2n$ corresponds to $A_1$-XXX spin chain with length $n$~\cite{Gorsky:1996hs}.
In this case, the gauge and flavor symmetries {are unrelated to} the global symmetry of the integrable system.
Then it's natural to ask how to obtain an integrable system possessing more general global symmetry groups.
It has been shown that global symmetry of the integrable system is characterized by the quiver structure of gauge theory~\cite{Nekrasov:2012xe,Nekrasov:2013xda}:
4d $\cN=2$ $\Gamma$-quiver gauge theory leads to $\Gamma$-XXX quantum spin chain model under the NS correspondence.
\paragraph{}
In order to see the correspondence between gauge theory and quantum integrable system, only one of the two equivariant parameters is turned on and identified with $\hbar$.
From this point of view, one may ask what happens on the moduli space of gauge theory if turning on both deformation parameters.
In such a {\em doubly quantum} situation, the quantum integrable system is promoted to the W-algebra, involving {\em noncommutative Hamiltonians}, and its algebraic structure is associated with the quiver structure.
This quiver gauge theoretic construction is called the quiver W-algebra~\cite{Kimura:2015rgi}, which actually turns out to be a dual to the AGT relation for gauge theory and conformal field theory~\cite{Alday:2009aq}, and its $q$-deformation~\cite{Awata:2009ur}. 
\paragraph{}
In this formalism, starting with 5d $\cN=1$ $\Gamma$-quiver gauge theory on $\mathbb{R}^4 \times S^1$, one obtains the algebra $W_{q_1,q_2}(\Gamma)$, which reproduces Frenkel--Reshetikhin's construction of $q$-deformation of W-algebra~\cite{Frenkel:1997}.
From 6d $\cN = (1,0)$ theory on $\mathbb{R}^4 \times T^2$, we obtain the elliptic deformation of W-algebra~\cite{Kimura:2016dys}.
Here we use the multiplicative equivariant parameters $(q_1,q_2) = (e^{\epsilon_1}, e^{\epsilon_2})$, which are the Cartan elements of $SO(4)$, spacetime rotation symmetry of $\mathbb{R}^4$.
Indeed, in the NS limit, $q_{1} \to 1$ or $q_2 \to 1$, all the non-commutative currents of W-algebra are reduced to be commutative, so that it describes infinitely many commuting conserved Hamiltonians of the quantum integrable system.
\paragraph{}
One obstacle of the correspondence between quiver gauge theory and integrable system/W-algebra is that the quiver structure is restricted to the so-called simply laced type quiver.
In other words, it has not been known how to construct the non-simply laced quiver from geometric point of view.
Recently there has been a construction of non-simply laced quiver (fractional quiver) which is based on the algebraic point of view, and reproduces Frenkel--Reshetikhin's $q$-deformation of W-algebra for $\Gamma \neq ADE$~\cite{Kimura:2017hez}.
An interesting specific feature of non-simply laced quiver is that the two equivariant parameters are not equalfooting anymore.
This implies that there are two possible classical limits of W-algebra, namely two distinct NS limits: $q_1 \to 1$ and $q_2 \to 1$.
\paragraph{}
In this paper, we study the two distinct NS limits of non-simply laced quiver gauge theory.
In particular, we will use the $BC_2$ quiver as the simplest nontrivial example of non-simply laced quiver to illustrate our general construction.
We find that the NS$_2$ limit $(q_2 \to 1)$ gives rise to $BC_2$ quantum spin chain, which is a natural generalization of~\cite{Nekrasov:2012xe,Nekrasov:2013xda} to non-simply laced quiver, while the NS$_1$ limit $(q_1 \to 1)$ leads to a twisted (degenerated) quantum integrable system obtained through the {\em naive} folding trick from $AD_3$ quiver, which is simply laced.
We will show this statement by exploring the asymptotic behavior of the gauge theory partition function in the NS$_{1,2}$ limits, namely the effective twisted superpotential $\msW_\text{eff}$, and the twisted $F$-term condition associated with $\msW_\text{eff}$ leads to the Bethe equation of the corresponding quantum integrable system.
In addition, the generating currents of W-algebra, constructed using the $qq$-character~\cite{Nekrasov:2015wsu},%
\footnote{See also~\cite{Nekrasov:2016qym,Nekrasov:2016ydq,Nekrasov:2017rqy,Nekrasov:2017gzb} and \cite{Bourgine:2015szm,Bourgine:2016vsq}.}
is reduced to the commutative currents in the classical limits, interpreted as the transfer matrix of the quantum integrable system.
\paragraph{}
This paper is organized as follows:
In Sec.~\ref{sec:gauge_th} we start with definitions of gauge theory partition function to fix our convention.
In particular we provide two expressions for the partition function, which are convenient to study the asymptotic limits.
In Sec.~\ref{sec:asymp} we study the two asymptotic limits of the partition function, namely the NS$_{1}$ limit $(q_1 \to 1)$ and the NS$_2$ limit $(q_2 \to 1)$.
In Sec.~\ref{sec:qq-ch} we consider the doubly quantum Seiberg--Witten curve, called the $qq$-character, for the non-simply laced quiver, and its behavior in the NS$_{1,2}$ limits.
We discuss its connection with the $q$-character, interpreted as the transfer matrix of the quantum integrable system.
In Sec.~\ref{sec:vac} we perform the saddle point analysis of the partition function in the NS$_{1,2}$ limits, and show its relation to the Bethe equation of the corresponding quantum integrable system.
We also discuss the root of Higgs branch of the moduli space to see the truncation of the infinitely many Bethe roots.
While we mostly focus on five dimensional non-simply laced quiver gauge theories in our main text, almost parallel analysis can be carried out also for their six dimensional counterparts, we collect some useful results for these in a separate Appendix~\ref{sec:6d}.


 \section{Gauge theory definitions}\label{sec:gauge_th}
\paragraph{}
 The partition function of four dimensional $\cN = 2$ supersymmetric quiver gauge theory defined using the path integral formalism turns out to be computable using the localization techniques
 (See \cite{ Pestun:2016zxk} for a detailed review):
 The infinite dimensional path integral is localized to finite {dimensional} contributions from the fixed points under the torus action in the instanton moduli space~\cite{Moore:1997dj,Nekrasov:2002qd}.
 Each {construction} can be written using the Chern characters associated with the torus actions. 
 {Here we shall introduce the relevant expressions for the partition functions using their Chern character formulae instead of the localization techniques which rely on the availability of the explicit Lagrangian, the benefit of our approach is that it is applicable even for non-Lagrangian theories, and application to four, five and six-dimensional quiver gauge theories with eight supercharges,}
\paragraph{}
 We begin by introducing the notations for the quiver diagrams, which specify the quiver gauge theories with eight supercharges in four (also five and six) dimensions to be considered, {again our construction here applies to both simply and non-simply laced quivers.}
 Let $\Gamma$ be a quiver, which consists of nodes $i \in \Gamma_0$ and edges $e \in \Gamma_1$, and the gauge group $U(n_i)$ is {assigned} to the $i$-th node.
 We also assign the ``root'' parameter to each quiver node, $(d_i \in \bZ_{>0})_{i \in \Gamma_0}$~\cite{Kimura:2017hez}.
 The fixed points in the instanton moduli space are labeled by a set of partitions $(\lambda_{i,\alpha})_{i \in \Gamma_0, \alpha \in [1,\ldots, n_i]}$ obeying the non-increasing condition, $\lambda_{i,\alpha,1} \ge \lambda_{i,\alpha,2} \ge \cdots \ge 0$ for $\forall(i,\alpha)$.
 The Chern characters are written using the equivariant parameters associated with the gauge and spacetime symmetries
 \begin{align}
  \nu_{i,\alpha} = e^{a_{i,\alpha}}
  \, , \qquad
  (q_1,q_2) = (e^{\epsilon_1}, e^{\epsilon_2})
  \, ,
 \end{align}
where $(a_{i,\alpha})_{i \in \Gamma_0, \alpha \in [1,\ldots,n_i]}$ is the Cartan part of the $i$-th gauge field, 
called the Coulomb moduli, and $(\epsilon_{1}, \epsilon_2)$ {are} the $\Omega$-background {parameters}.
 We can also consider the mass parameters for (anti)fundamental matters as the equivariant parameters for the global flavor symmetry.
 The Chern characters for the corresponding bundles are given as follows:
 \begin{align}
  \bN_i = \sum_{\alpha=1}^{n_i} \nu_{i,\alpha}
  \, , \qquad
  \bK_i = \sum_{\alpha=1}^{n_i} \sum_{(s_1,s_2) \in \lambda_{i,\alpha}}
  q_1^{d_i(s_1 - 1)} q_2^{s_2 - 1} \nu_{i,\alpha}
  \, .
 \end{align}
 We remark that their dimensions are given by the rank of the gauge group and the instanton number for the $i$-th node, respectively:
\begin{align}
 \operatorname{dim}_{\bC} \bN_i = n_i
 \, , \qquad
 \operatorname{dim}_{\bC} \bK_i = \sum_{\alpha=1}^{n_i} \sum_{k=1}^\infty |\lambda_{i,\alpha,k}| = k_i
 \, ,
\end{align}
where we use the same notation for the bundle itself as the corresponding character as long as no confusion.
Then the universal bundle for the $i$-th node is constructed from these bundles,
\begin{align}
 \bY_i = \bN_i - \left( \bigwedge \bQ_i \right) \bK_i
 \, ,
\end{align}
where we define $\bigwedge \bQ_i = (1 - q_1^{d_i})(1 - q_2)$ associated with the spacetime {rotation} symmetry.
Although it is symmetric under exchange $(q_1^{d_i} \leftrightarrow q_2)$, the corresponding character has seemingly two different expressions:
\begin{subequations} 
 \begin{align}
  \bY_i
  & =
  (1 - q_1^{d_i}) \sum_{\alpha=1}^{n_i} \sum_{k=1}^\infty
  q_1^{d_i(k-1)} q_2^{\lambda_{i,\alpha,k}} \nu_{i,\alpha}
  =: (1 - q_1^{d_i}) \sum_{x \in \cX_i} x,
  \\
  & =
  (1 - q_2) \sum_{\alpha=1}^{n_i} \sum_{k=1}^\infty
  q_1^{d_i \lambda_{i,\alpha,k}^\text{T}} q_2^{k-1} \nu_{i,\alpha}
  =: (1 - q_2) \sum_{\tilde{x} \in {\cX}_i^\text{T}} \tilde{x},
 \end{align} 
\end{subequations}
with the set of configurations
\begin{subequations}
\begin{align}
 \mathcal{X}_i = (x_{i,\alpha,k})_{i \in \Gamma_0,\, \alpha \in [1,\ldots,n_i],\, k \in [1,\ldots,\infty]}
 \, , \qquad
 \cX = \bigsqcup_{i \in \Gamma_0} \cX_i
 \, ,
\end{align}
\begin{align}
 {\mathcal{X}}^\text{T}_i = (\tilde{x}_{i,\alpha,k})_{i \in \Gamma_0,\, \alpha \in [1,\ldots,n_i],\, k \in [1,\ldots,\infty]} 
 \, , \qquad
 {\cX}^\text{T} = \bigsqcup_{i \in \Gamma_0} \cX_i^\text{T}
 \, ,
\end{align}
\end{subequations}
where the dynamical variables are defined
\begin{align}
 x_{i,\alpha,k} =
 q_2^{\lambda_{i,\alpha,k}} q_1^{d_i(k-1)} \nu_{i,\alpha}
 \, , \qquad
 \tilde{x}_{i,\alpha,k} =
 q_1^{d_i\lambda_{i,\alpha,k}^\text{T}} q_2^{k-1} \nu_{i,\alpha}
 \, ,
\end{align}
and $(\lambda_{i,\alpha}^\text{T})_{i \in \Gamma_0, \alpha \in [1,\ldots,n_i]}$ denotes the transposed partition.
We can view assigning the root $d_i$ to each gauge node as rescaling the columns or rows of the corresponding Young diagram labeling the partition by a scale factor $d_i$.
We remark that $\cX$ and ${\cX}^\text{T}$ are isomorphic to each other since both sets take values from the same Young diagrams $(\lambda_{i,\alpha,k})$ and their transposition $(\lambda_{i, \alpha, k}^{\text{T}})$. 
\paragraph{}
We can compute the Chern characters for the vector and hyper-multiplet contributions from the universal bundle:
\begin{subequations} 
 \begin{align}
 \bV_i
  & =
  \frac{\bY_i^\vee \bY_i}{\bigwedge \bQ_i}
  =
   \begin{cases}
    \displaystyle
    \frac{1 - q_1^{-d_i}}{1 - q_2}
    \sum_{(x,x') \in \cX_i \times \cX_i} \frac{x'}{x}
    \\
    \displaystyle
    \frac{1 - q_2^{-1}}{1 - q_1^{d_i}}
    \sum_{(\tx,\tx') \in {\cX}^\text{T}_i \times {\cX}^\text{T}_i} \frac{\tx'}{\tx}
   \end{cases}
 \label{eq:V_ch} \\[.5em]
 \bH_{e:i \to j}
  & =
  - \bM_e \frac{\bY_i^\vee \bY_j}{\bigwedge \bQ_{ij}}
  =
  \begin{cases}
   \displaystyle
   - \mu_{e} \frac{(1 - q_1^{-d_i})(1 - q_1^{d_j})}{(1 - q_2)(1 - q_1^{d_{ij}})}
   \sum_{(x,x') \in \cX_i \times \cX_j} \frac{x'}{x}
   \\ \displaystyle
   - \mu_{e} \frac{1 - q_2^{-1}}{1 - q_1^{d_{ij}}}
   \sum_{(\tx,\tx') \in {\cX}^\text{T}_i \times {\cX}^\text{T}_j} \frac{\tx'}{\tx}      
  \end{cases}
  \label{eq:H_ch}
 \end{align}
\end{subequations}
where we define $\bigwedge \bQ_{ij} = (1 - q_1^{d_{ij}})(1 - q_2)$ with $d_{ij} = \operatorname{gcd}(d_i,d_j)$, and $\bM_{e} = \mu_e$ for $e \in \Gamma_1$ denotes the multiplicative mass parameter for the bifundamental hypermultiplet assigned to the edge $e \in \Gamma_1$.
We define the dual character $\bX^\vee = \sum x^{-1}$ with respect to the character $\bX = \sum x$,
again we have two expressions for them with respect to $(\lambda_{i,\alpha})$ and $(\lambda_{i,\alpha}^\text{T})$.
These contributions are nicely combined into a simple form:
\begin{subequations}
\begin{align}
 \sum_{i \in \Gamma_0} \bV_i + \sum_{e:i \to j} \bH_{e:i \to j}
 & =
 \sum_{(x,x') \in \cX \times \cX}
 \left( c^+_{\si(x) \si(x')}\right)
 \frac{1 - q_1^{d_{\si(x')}}}{1 - q_2}
 \frac{x'}{x},
 \label{eq:VandH1}
 \\
 & =
 \sum_{(\tx,\tx') \in \cX^\text{T} \times \cX^\text{T}}
 \left( c^+_{\si(\tx) \si(\tx')}\right)^\vee
 \frac{1 - q_2^{-1}}{1 - q_1^{d_{\si(\tx)}}}
 \frac{\tx'}{\tx},
 \label{eq:VandH2} 
\end{align}
\end{subequations}
where $\si: \cX \to \Gamma_0$ is the node label $\si(x) = i$ if $x \in \cX_i$ (or if $\tx \in \cX_i^{\text{T}}$), and we define the mass-deformed (an {upper} half of) Cartan matrix:%
\footnote{%
In this paper we don't use the other half of the Cartan matrix which is necessary to discuss the relation to the quiver W-algebra formalism.
See~\cite{Kimura:2017hez} for details.
}
\begin{align}
 c_{ij}^+
 & =
 \delta_{ij} - \sum_{e:i \to j} \mu_e^{-1}
 \frac{1 - q_1^{-d_i}}{1 - q_1^{-d_{ij}}}
 =
 \delta_{ij} - \sum_{e:i \to j} \sum_{r=0}^{d_i/d_{ij}-1}
 \mu_e^{-1} q_1^{-rd_{ij}}
 \, .
 \label{eq:Cartan}
\end{align}
The last expression implies the duplication of the bifundamental contribution depends on the root parameters $\{d_i\}$ and $\{d_{ij}\}$ as:
$\{\mu_e\} \to \{\mu_e q_1^{r d_{ij}}\}_{r \in [0,\ldots,d_i/d_{ij}-1]} = \{\mu_e,\mu_e q_1^{d_{ij}},\ldots,\mu_e q_1^{d_i-d_{ij}}\}$.
We remark that swapping $q_1^{d_i} \leftrightarrow q_2$ corresponds to transposition of the Cartan matrix $c_{ij}^+ \leftrightarrow c_{ji}^+$.
In the classical limit, this Cartan matrix is reduced as
\begin{align}
 c_{ij}^+
 \ \longrightarrow \
 \delta_{ij} - \#(e: i \to j)
 \label{eq:Cartan_cl} 
\end{align}
where the number of edges is counted with the multiplicity $d_{i}/d_{ij}$:
\begin{align}
 \#(e: i \to j) =
 \sum_{e:i \to j} \frac{d_i}{d_{ij}}
 \, .
 \label{eq:edge_num}  
\end{align}
Furthermore the expression \eqref{eq:VandH1} becomes simpler by introducing the symmetrized Cartan matrix
\begin{align}
 b_{ij}^+ := \frac{1 - q_1^{d_j}}{1 - q_1} c_{ij}^+
 = \delta_{ij} - \sum_{e:i \to j} \mu_e^{-1}
 \frac{1 - q_1^{-d_i}}{1 - q_1}
 \frac{1 - q_1^{d_j}}{1 - q_1^{-d_{ij}}}
\end{align}
as follows,
\begin{align}
 \sum_{i \in \Gamma_0} \bV_i + \sum_{e:i \to j} \bH_{e:i \to j}
 & =
 \sum_{(x,x') \in \cX \times \cX}
 \left( b^+_{\si(x) \si(x')}\right)
 \frac{1 - q_1}{1 - q_2}
 \frac{x'}{x}
 \, ,
\end{align}
while we observe that such a simplification does not occur for the other expression \eqref{eq:VandH2}.
 \paragraph{}
In addition to the vector and bifundamental hypermultiplets, we can consider the fundamental and antifundamental matters, which are obtained from the bifundamental factor through the following reductions
\begin{align}
 \bY_i \ \longrightarrow \
 \bM_i
 = \sum_{f=1}^{n_i^\text{f}} \mu_{i,f}
 =: \sum_{\mu \in \cX_i^\text{f}} \mu
 \, , \qquad
 \bY_i^\vee \ \longrightarrow \
 \widetilde{\bM}_i
 = \sum_{f=1}^{n_i^\text{af}} \tilde{\mu}_{i,f}
 =: \sum_{\mu \in \cX_i^\text{af}} \mu
\end{align}
where we define sets of the (anti)fundamental mass parameters,
\begin{align}
 \cX_i^\text{f}
 = (\mu_{i,f})_{i \in \Gamma_0, \, f \in [1,\ldots,n_i^\text{f}]}
 \, , \qquad
 \cX_i^\text{af}
 = (\tilde{\mu}_{i,f})_{i \in \Gamma_0, \, f \in [1,\ldots,n_i^\text{af}]}
 \, .
\end{align}
The corresponding Chern characters are given by
\begin{align}
 \bH_i^\text{f}
 =
 - \frac{\bY_i^\vee \bM_i}{\bigwedge \bQ_i}
 =
 \begin{cases}
  \displaystyle
  \frac{q_1^{-d_i}}{1 - q_2}
  \sum_{(x,\mu) \in \cX_i \times \cX_i^\text{f}} \frac{\mu}{x}
  \\
  \displaystyle
  \frac{q_2^{-1}}{1 - q_1^{d_i}}
  \sum_{(\tx,\mu) \in \cX^\text{T}_i \times \cX_i^\text{f}} \frac{\mu}{\tx}  
 \end{cases}
 \quad
 \bH_i^\text{af}
 =
 - \frac{\widetilde{\bM}_i \bY_i}{\bigwedge \bQ_i}
 =
 \begin{cases}
  \displaystyle
  - \frac{1}{1 - q_2}
  \sum_{(\mu,x) \in \cX_i^\text{af} \times \cX_i} \frac{x}{\mu}
  \\
  \displaystyle
  - \frac{1}{1 - q_1^{d_i}}
  \sum_{(\mu,\tx) \in \cX_i^\text{af} \times \cX^\text{T}_i} \frac{\tx}{\mu}  
 \end{cases}
\end{align}
Here we observe that in contrast with the bifundamental hypermultiplets, we again have the symmetry under the $(q_1^{d_i}, \cX_i^{\text{T}})\leftrightarrow (q_2, {\cX}_i) $ exchange.
\paragraph{}
Similarly, the partition function ($Z$-function) for a 5d or 6d quiver gauge theory with eight supercharges compactified on a circle $\bR^4 \times S^1$ or a torus $\bR^4 \times T^2$ 
can also be obtained from the corresponding characters by applying an {index-like computation}, also called the plethystic exponential, defined as \cite{Kimura:2016dys}:
\begin{align}
 \text{(5d):} \quad
 \ind \left[ \sum_{k} x_k \right]
 = \prod_{k} (1 - x_k^{-1})
 \, , \qquad
 \text{(6d):} \quad
 \ind_p \left[ \sum_{k} x_k \right]
 = \prod_k \theta(x_k^{-1};p)
 \label{eq:index}
\end{align}
where $p = \exp \left( 2\pi \iota \tau\right)$ 
is the multiplicative modulus of the torus with $\iota = \sqrt{-1}$, and the theta function is defined through the $q$-Pochhammer symbol $(x; q)_n = \prod_{k=0}^{n-1}(1-x q^k)$ as:
\begin{align}
 \theta(x;p) = (x;p)_\infty (px^{-1};p)_\infty
 \, .
 \label{eq:th_fn}
\end{align}
We remark that these two indices obey the same reflection relation:
\begin{align}
 \ind \left[ \bX^\vee \right]
 = (-1)^{\operatorname{rk} \bX} (\det \bX) \, \ind \left[ \bX \right]
 \, ,
\end{align}
and the 4d partition function is obtained by the corresponding Chern class as a reduction from 5d/6d theory.
Here we focus on the explicit expressions for the partition functions for the five dimensional quiver gauge theories, we relegate almost parallel discussions for six dimensional theories into Appendix~\ref{sec:6d}:
\begin{subequations}
\begin{align}
 Z_i^\text{vec} = \ind \left[ \bV_i \right]
 & =
 \prod_{(x,x') \in \cX_i \times \cX_i}
 \left(q_1^{d_i} q_2 \frac{x}{x'}; q_2\right)_\infty
 \left(q_2 \frac{x}{x'}; q_2\right)_\infty^{-1}\,, 
 \\
 & =
 \prod_{(\tx,\tx') \in \cX^\text{T}_i \times \cX^\text{T}_i}
 \left(q_1^{d_i} q_2 \frac{\tx}{\tx'}; q_1^{d_i}\right)_\infty
 \left(q_1^{d_i} \frac{\tx}{\tx'}; q_1^{d_i}\right)_\infty^{-1}
 \, , 
\end{align}
\begin{align}
 Z_{e:i \to j}^\text{bf}
 = \ind \left[ \bH_{e:i \to j}^\text{bf} \right]
 & =
 \prod_{(x,x') \in \cX_i \times \cX_j}
 \prod_{r=0}^{d_i/d_{ij}-1}
 \left(\mu_e^{-1} q_1^{d_i-rd_{ij}} q_2 \frac{x}{x'}; q_2\right)_\infty^{-1}
 \left(\mu_e^{-1} q_1^{-rd_{ij}} q_2 \frac{x}{x'}; q_2\right)_\infty\,,
 \\
 & =
 \prod_{(\tx,\tx') \in \cX^\text{T}_i \times \cX^\text{T}_j}
 \prod_{r=0}^{d_i/d_{ij}-1}
 \left(\mu_e^{-1} q_1^{d_j-rd_{ij}} q_2 \frac{\tx}{\tx'}; q_1^{d_j}\right)_\infty^{-1}
 \left(\mu_e^{-1} q_1^{d_j-rd_{ij}} \frac{\tx}{\tx'}; q_1^{d_j}\right)_\infty
 \, ,
\end{align}
\begin{align} 
 Z_i^\text{f}
 = \ind \left[ \bH_i^\text{f} \right]
 & =
 \prod_{(x,\mu) \in \cX_i \times \cX_i^\text{f}}
 \left( q_1^{d_i} q_2 \frac{x}{\mu}; q_2 \right)_{\infty}^{-1}\,,
 =
 \prod_{(\tx,\mu) \in \cX^\text{T}_i \times \cX_i^\text{f}}
 \left( q_1^{d_i} q_2 \frac{\tx}{\mu}; q_1^{d_i} \right)_{\infty}^{-1}
 \, , \label{eq:Zf_5d} \\[.5em]
 Z_i^\text{af}
 = \ind \left[ \bH_i^\text{af} \right]
 & =
 \prod_{(\mu,x) \in \cX_i^\text{af} \times \cX_i}
 \left( q_2 \frac{\mu}{x}; q_2 \right)_{\infty}
 =
 \prod_{(\mu,\tx) \in \cX_i^\text{af} \times \cX^\text{T}_i}
 \left( q_1^{d_i} \frac{\mu}{\tx}; q_1^{d_i} \right)_{\infty}\,.
 \end{align}
\end{subequations}
These contributions computed using Chern characters correspond to the full partition function containing both the perturbative and non-perturbative instanton factors.
The perturbative part is similarly formulated with the ground state configuration
\begin{subequations}
\begin{align}
 \mathring{\cX}_i = (\mathring{x}_{i,\alpha,k})_{i \in \Gamma_0,\, \alpha \in [1,\ldots,n_i],\, k \in [1,\ldots,\infty]}
 \, , \qquad
 &
 \mathring{\cX} = \bigsqcup_{i \in \Gamma_0} \cX_i
 \, , \\[.5em]
 \mathring{{\mathcal{X}}}^\text{T}_i = (\mathring{\tilde{x}}_{i,\alpha,k})_{i \in \Gamma_0,\, \alpha \in [1,\ldots,n_i],\, k \in [1,\ldots,\infty]}
 \, , \qquad
 &
 \mathring{{\cX}}^\text{T} = \bigsqcup_{i \in \Gamma_0} \mathring{{\cX}}^\text{T}_i
\end{align}
\end{subequations}
which correspond to the empty Young diagrams $(\lambda_{i,\alpha} = \emptyset)$,
\begin{align}
 \mathring{x}_{i,\alpha,k} = q_1^{d_i(k-1)} \nu_{i,\alpha}
 \, , \qquad
 \mathring{\tilde{x}}_{i,\alpha,k} = q_2^{(k-1)} \nu_{i,\alpha}
 \, .
\end{align}
The perturbative partition functions are given by the index evaluated with the {following background configurations}:
\begin{align}
 \mathring{Z}_i^\text{vec} = \ind \left[ \bV_i \right]\Big|_{\cX \to \mathring{\cX}}
 \, , \qquad
 \mathring{Z}_{e:i \to j}^\text{bf} = \ind \left[ \bH_{e:i \to j} \right]\Big|_{\cX \to \mathring{\cX}}
 \, .
\end{align}
The same result is obtained by the replacement $\cX^\text{T} \to \mathring{\cX}^\text{T}$.
Then the instanton part is given by the ratios of the full and perturbative contributions:
\begin{align}
 Z_i^\text{vec,inst}
 =
 Z_i^\text{vec} / \mathring{Z}_i^\text{vec}
 \, , \qquad
 Z_{e:i \to j}^\text{bf,inst}
 =
 Z_{e:i \to j}^\text{bf} / \mathring{Z}_{e:i \to j}^\text{bf}
 \, .
\end{align}
In addition, we have the topological factor corresponding to the instanton contributions:
\begin{subequations}
\begin{align}
 Z_i^\text{top}
 =
 \frkq_i^{|\lambda_{i}|}
 & =
 \exp
 \left(
 \log \frkq_i
 \left(
 \sum_{x \in \cX_i} \log_{q_2} x
 - \sum_{\mathring{x} \in \mathring{\cX}_i} \log_{q_2} \mathring{x}
 \right)
 \right)
 \\
 & =
 \exp
 \left(
 \frac{1}{d_i}
 \log \frkq_i
 \left(
 \sum_{\tx \in {\cX}^\text{T}_i} \log_{q_1} \tx
 - \sum_{\mathring{\tx} \in {\mathring{\cX}}^\text{T}_i} \log_{q_1} \mathring{\tx}
 \right)
 \right) 
\end{align} 
\end{subequations}
where $\frkq_i = e^{2\pi \iota \tau_i}$ is the multiplicative {gauge} coupling constant for the {gauge} node $i \in \Gamma_0$, and $|\lambda_i| = \sum_{\alpha=1}^{n_i} \sum_{k=1}^\infty \lambda_{i,\alpha,k} = \sum_{\alpha=1}^{n_i} \sum_{k=1}^\infty \lambda_{i,\alpha,k}^\text{T}$.
\paragraph{}
Finally, all these contributions are combined to give the total partition function
which sums over all the possible configurations%
\footnote{Here we have ignored the Chern--Simons term contributions to the five dimensional partition functions.}:
\begin{align}
 Z = \sum_{\cX} Z^\text{tot}_\cX
 \label{eq:Z_sum}
\end{align}
with all the vector and hypermultiplet contributions for a given $\cX$ packaged into:
\begin{align}
 Z^\text{tot}_\cX
 =
 \prod_{i \in \Gamma_0}
  Z_i^\text{vec} Z_i^\text{f} Z_i^\text{af} Z_i^\text{top} 
 \prod_{e \in \Gamma_1}
  Z_e^\text{bf} \,
 \Bigg|_{\cX}
 \, .
 \label{eq:Z_tot}
\end{align}
For our later purpose, it is also convenient to split the total partition function into perturbative and topological non-perturbative parts as:
\begin{eqnarray}
Z &=& \left(\prod_{i\in \Gamma_0} \mathring{Z}_i^{\text{vec}} \mathring{Z}_i^{\text{f}}  \mathring{Z}_i^{\text{af}}   \prod_{e\in \Gamma_1} \mathring{Z}^{\text{bf}}_e \right)
\sum_{\cX}   \exp\left(\log\Xi_{\cX}(x) - \log \Xi_{\mathring{\cX}}(\mathring{x}) \right)\nn\\
&=& \left(\prod_{i\in \Gamma_0} \mathring{Z}_i^{\text{vec}} \mathring{Z}_i^{\text{f}}  \mathring{Z}_i^{\text{af}}   \prod_{e\in \Gamma_1} \mathring{Z}^{\text{bf}}_e \right)
\sum_{\cX^{\text{T}}}   \exp\left(\log\Xi_{\cX^{\text{T}}}(\tx) - \log \Xi_{\mathring{\cX}^{\text{T}}}(\mathring{\tx}) \right)
\end{eqnarray}
where 
\bea
\log \Xi_{\cX}(x) &=& \sum_{i\in  \Gamma_0} \Big\{
\sum_{(\alpha,k) \neq (\alpha',k')}
\log \frac{\left(q_1^{d_i}q_2 \frac{x_{i,\alpha, k}}{x_{i, \alpha', k'}}; q_2\right)_{\infty}}{\left(q_1^{d_i} \frac{x_{i,\alpha, k}}{x_{i, \alpha', k'}}; q_2\right)_{\infty}}
+\frac{2\pi i \tau_i}{\epsilon_2} \sum_{\alpha=1}^{n_i} \sum_{k=1}^\infty \log x_{i, \alpha, k}
\nonumber\\
&+&\sum_{\alpha=1}^{n_i} \sum_{k=1}^\infty \sum_{\tmu \in \cX_i^{\text{af}}}\log \left(q_2\frac{\tmu}{x_{i, \alpha, k}}; q_2\right)_\infty
-\sum_{\alpha=1}^{n_i} \sum_{k=1}^\infty \sum_{\mu \in \cX_i^{\text{f}}}\log \left(q_1^{d_i} q_2\frac{x_{i, \alpha, k}}{\mu}; q_2\right)_\infty\Big\}\nonumber\\
&+& \sum_{e\in \Gamma_1}
\sum_{\alpha=1}^{n_i} \sum_{\alpha'=1}^{n_j} \sum_{k=1}^{\infty} \sum_{k'=1}^{\infty} \sum_{r = 0}^{d_i/d_{ij}-1} \log \frac{ \left(\frac{q_2 q_1^{-r d_{ij}}}{\mu_e} \frac{x_{i,\alpha, k}}{x_{j, \alpha', k'}}; q_2\right)_\infty }{\left(\frac{q_2 q_1^{d_j-r d_{ij}} }{\mu_e} \frac{x_{i,\alpha, k}}{x_{j, \alpha', k'}}; q_2\right)_\infty }.\\
 \log \Xi_{\cX^\text{T}}(\tx) 
 &=& \sum_{i\in  \Gamma_0} \Big\{
 \sum_{(\alpha,k) \neq (\alpha',k')} 
 \log \frac{\left(q_1^{d_i}q_2 \frac{\tx_{i,\alpha, k}}{\tx_{i, \alpha', k'}}; q_1^{d_i}\right)_{\infty}}{\left(q_1^{d_i} \frac{\tx_{i,\alpha, k}}{\tx_{i, \alpha', k'}}; q_1^{d_i}\right)_{\infty}}
 +\frac{2\pi i \tau_i}{d_i \epsilon_1} \sum_{\alpha=1}^{n_i} \sum_{k=1}^\infty \log \tx_{i, \alpha, k} 
\nonumber\\
&+&\sum_{\alpha=1}^{n_i} \sum_{k=1}^\infty \sum_{\tmu \in \cX_i^{\text{af}}}\log \left(q_1^{d_i}\frac{\tmu}{\tx_{i, \alpha, k}}; q_1^{d_i}\right)_\infty
 -\sum_{\alpha=1}^{n_i} \sum_{k=1}^\infty \sum_{\mu \in \cX_i^{\text{f}}}\log \left(q_1^{d_i} q_2\frac{\tx_{i, \alpha, k}}{\mu}; q_1^{d_i}\right)_\infty\Big\}\nonumber\\
&+& \sum_{e\in \Gamma_1} \sum_{\alpha=1}^{n_i} \sum_{\alpha'=1}^{n_j} \sum_{k=1}^{\infty} \sum_{k'=1}^{\infty} \sum_{r = 0}^{d_i/d_{ij}-1} \log \frac{ \left(\frac{q_1^{d_j-r d_{ij}}}{\mu_e} \frac{\tx_{i,\alpha, k}}{\tx_{j, \alpha', k'}}; q_1^{d_j}\right)_\infty }{\left(\frac{q_1^{d_j-r d_{ij}} q_2 }{\mu_e} \frac{\tx_{i,\alpha, k}}{\tx_{j, \alpha', k'}}; q_1^{d_j}\right)_\infty }
\eea
and $\log \Xi_{\mathring{\cX}(\mathring{x})}$ and $\log \Xi_{\mathring{\cX}^\text{T}}(\mathring{\tilde{x}})$ are given by replacing $(x_{i,\alpha, k}, \cX)$ and $(\tx_{i,\alpha, k}, \cX^{\text{T}})$ with $(\mathring{x}_{i, \alpha, k}, \mathring{\cX})$ and  $(\mathring{\tx}_{i, \alpha, k}, \mathring{\cX}^{\text{T}})$.

\subsection{$BC_2$ quiver}
\paragraph{}
Let us explicitly write down the partition function for $BC_2$ quiver with $(d_1,d_2) = (2,1)$ and $d_{12} = 1$, 
this will serve as our simplest illustrative example.
In particular the bifundamental contribution for 5d partition function is given as follows:
\begin{subequations}
\begin{align}
 Z_{1 \to 2}^\text{bf}
 & =
 \prod_{(x,x') \in \cX_2 \times \cX_1}
 \left(
  \mu q_1 \frac{x}{x'};q_2
 \right)_\infty^{-1}
 \left(
  \mu q_1^{-1} \frac{x}{x'};q_2
 \right)_\infty
 \left(
  \mu \frac{x}{x'};q_2
 \right)_\infty^{-1}
 \left(
  \mu q_1^{-2} \frac{x}{x'};q_2
 \right)_\infty   
 \\
 & =
 \prod_{(x,x') \in \cX^\text{T}_2 \times \cX^\text{T}_1}
 \left(
  \mu \frac{x}{x'};q_1^2
 \right)_\infty^{-1}
 \left(
  \mu q_2^{-1} \frac{x}{x'};q_1^2
 \right)_\infty
 \left(
  \mu q_1^{-1} \frac{x}{x'};q_1^2
 \right)_\infty^{-1}
 \left(
  \mu q_1^{-1} q_2^{-1} \frac{x}{x'};q_1^2
 \right)_\infty  
 \\
 Z_{2 \to 1}^\text{bf}
 & =
 \prod_{(x,x') \in \cX_1 \times \cX_2}
 \left(
  \mu^{-1} q_1^2 q_2 \frac{x}{x'};q_2
 \right)_\infty^{-1}
 \left(
  \mu^{-1} q_2 \frac{x}{x'};q_2
 \right)_\infty
 \\
 & =
 \prod_{(x,x') \in \tilde\cX_1 \times \tilde\cX_2}
 \left(
  \mu^{-1} q_1 q_2 \frac{x}{x'};q_1
 \right)_\infty^{-1}
 \left(
  \mu^{-1} q_1 \frac{x}{x'};q_1
 \right)_\infty
\end{align}
\end{subequations}
where $\mu = \mu_{2 \to 1} = \mu_{1 \to 2}^{-1} q_1 q_2$.
The contribution of the edge $e:1 \to 2$ is duplicated with shifted mass parameters $\{\mu_{1 \to 2}, \mu_{1 \to 2} q_1\}$, 
while there is a single contribution of the edge $2 \to 1$ as usual.
The remaining contributions assigned to the node $i = 1, 2$, vector and (anti)fundamental contributions, are simply given by replacing $q_1 \to q_1^{d_{i=1,2}}$.

\section{Asymptotic limit and saddle point configurations}\label{sec:asymp}
\paragraph{}
In this section, we consider the asymptotic limits of the various partition functions reviewed earlier, 
in particular we will focus on the non-simply laced quiver gauge theories and use $BC_2$ quiver as a prototype example.
In general, the gauge theory partition functions computed in $\Omega$-background depend on the equivariant parameters $(\epsilon_1, \epsilon_2)$, in the limit $\epsilon_{1,2} \to 0$ \cite{Nekrasov:2009rc}, the critical saddle point configuration dominates in the summation \eqref{eq:Z_sum}.
In this particular limit, it was shown for various simply laced quiver gauge theories that the defining equations of saddle point configurations precisely coincide with the Bethe ansatz equations of the underlying quantum integrable systems, whose corresponding classical spectral curves can be identified with the Seiberg--Witten curves. 
For simply laced quivers, the limits $\epsilon_1 \to 0$ or $\epsilon_2 \to 0$ provide the identical result modulo trivial relabeling, hence the same quantum integrable systems emerge from the their partition functions.
Interestingly, for the non-simply laced quivers, as we will see momentarily, different quantum integrable systems can arise from each of the $\epsilon_1\to 0$ and $\epsilon_2 \to 0$ limit.
Since we expect the two $\Omega$-background parameters play different roles for non-simply laced quivers, 
we shall name the two asymptotic limits $\epsilon_1 \to 0$ (NS$_1$ limit) and $\epsilon_2 \to 0$ (NS$_2$ limit).
\paragraph{}
Schematically, the asymptotic behaviors of 5d/6d gauge theory partition functions take the generic form through the following twisted superpotentials:
\begin{align}
 \lim_{\epsilon_1 \to 0} \epsilon_1 \log Z_{\rm 5d/6d}(\epsilon_1, \epsilon_2) = \msW_1(\epsilon_2)
 \quad \text{and} \quad
 \lim_{\epsilon_2 \to 0} \epsilon_2 \log Z_{\rm 5d/6d}(\epsilon_1, \epsilon_2) = \msW_2(\epsilon_1)
 \,.
\end{align}
As we will summarize below, their explicit expressions can often be expressed in terms of polylogarithm functions and their elliptic generalizations, 
which also often occur in the low energy effective descriptions of the compactified three and four dimensional gauge theories with four supercharges.
Later we will discuss that this is not a mere coincidence, rather we can regard these as decoupling limits which restrict the remaining light degrees of freedoms 
along certain co-dimension two sub-manifolds. After imposing appropriate truncation conditions, these low energy theories can be potentially identified with the world volume theories of the co-dimension two topological defects.
\paragraph{}
Here we summarize the relevant mathematical formulae for considering the NS$_1$ and NS$_2$ limits for the compactified five dimensional non-simply laced quiver gauge theories. 
\subsection{Asymptotic limit}\label{sec:5d/3d}
\paragraph{}
For $q = e^\epsilon$, the quantum dilogarithm has the following small $\epsilon$-expansion:
\begin{align}
 (z;q)_\infty^{-1}
 & =
 \exp
 \left(
 \sum_{m=1}^\infty \frac{z^m}{m(1 - q^m)}
 \right)
 =
 \exp
 \left(
  - \frac{1}{\epsilon} \Li_2(z) + O(\epsilon^0)
 \right)
 \, ,
\end{align}
where we define the polylogarithm function
\begin{align}
 \Li_k(z) = \sum_{n=1}^\infty \frac{z^n}{n^k}
 \, .
\end{align}
The sub-leading terms have an explicit expression in terms of the Bernoulli numbers which are not required for the subsequent discussions.
We take the following combinations of the quantum dilogarithms which appears in the gauge theory partition functions:
\begin{subequations} 
  \begin{align}
   \frac{\left( q_1^{d_i} z; q_1^{d_i} \right)_\infty}
   {\left( q_1^{d_i} q_2 z; q_1^{d_i} \right)_\infty}
   & \stackrel{\epsilon_1 \to 0}{\longrightarrow} \
   \exp
   \left(
   - \frac{1}{d_i \epsilon_1} \left( \Li_2(q_2 z) - \Li_2(z) \right) 
   \right)
   =
   \exp \left( - \frac{1}{d_i \epsilon_1} L(z;q_2) \right)\,,
   \\[.5em] 
  \frac{\left( q_2 z; q_2 \right)_\infty}
   {\left( q_1^{d_i} q_2 z; q_2 \right)_\infty}
   & \stackrel{\epsilon_2 \to 0}{\longrightarrow} \
   \exp
 \left(
   - \frac{1}{\epsilon_2} \left( \Li_2(q_1^{d_i} z) - \Li_2(z) \right) 
   \right)
   =
 \exp \left( - \frac{1}{\epsilon_2} L(z;q_1^{d_i}) \right)\,,
  \end{align}
\end{subequations}
where we have defined:
\begin{align}
 L(z;q) = \Li_2(q z) - \Li_2(z)
 \, .
 \label{eq:L-func}
\end{align}
Thus we obtain the asymptotic behavior of the various contributions to the partition functions of non-simply laced quiver gauge theories in the NS$_{1,2}$ limits:
\begin{subequations}
\begin{align}
 Z_i^\text{vec}
 & \longrightarrow \
 \begin{cases}
  \displaystyle
  \exp
  \left(
  \frac{1}{d_i \epsilon_1}
  \sum_{(\tx,\tx') \in \cX^\text{T}_i \times \cX^\text{T}_i}
  L \left(\frac{\tx}{\tx'};q_2\right)
  \right)
  & (q_1 \to 1)\,,
  \\[1.5em]
  \displaystyle
  \exp
  \left(
  \frac{1}{\epsilon_2}
  \sum_{(x,x') \in \cX_i \times \cX_i}
  L \left(\frac{x}{x'};q_1^{d_i}\right)
  \right)
  & (q_2 \to 1)\,,
 \end{cases}
\end{align}
\begin{align}
 Z_{e:i \to j}^\text{bf}
 & \longrightarrow \
 \begin{cases}
  \displaystyle
  \exp
  \left(
  - \frac{1}{d_{ij} \epsilon_1}
  \sum_{(\tx,\tx') \in \cX^\text{T}_i \times \cX^\text{T}_j}  
  L \left( \mu_e^{-1} \frac{\tx}{\tx'};q_2 \right)
  \right)
  & (q_1 \to 1)
  \\[1.5em]
  \displaystyle
  \exp
  \left(
  - \frac{1}{\epsilon_2}
  \sum_{(x,x') \in \cX_i \times \cX_j}
  \sum_{r = 0}^{d_i/d_{ij} - 1}
  L \left( \mu_e^{-1} q_1^{-rd_{ij}} \frac{x}{x'};q_1^{d_j} \right)
  \right)
  & (q_2 \to 1)
 \end{cases}
\end{align}
\begin{align}
 Z_i^\text{f}
 & \longrightarrow \
 \begin{cases}
  \displaystyle
  \exp
  \left(
   - \frac{1}{d_i \epsilon_1} \sum_{(\tx,\mu) \in {\cX}^\text{T}_i \times \cX_i^\text{f}} \Li_2 \left( q_2 \frac{\tx}{\mu} \right)
  \right)
    & (q_1 \to 1)
  \\[1.5em]
  \displaystyle
  \exp
  \left(
   - \frac{1}{\epsilon_2} \sum_{(x,\mu) \in {\cX}_i \times \cX_i^\text{f}} \Li_2 \left( q_1^{d_i} \frac{x}{\mu} \right)
  \right)  
  & (q_2 \to 1)  
 \end{cases}
\end{align}
\begin{align}
 Z_i^\text{af}
 & \longrightarrow \
 \begin{cases}
  \displaystyle
  \exp
  \left(
   \frac{1}{d_i \epsilon_1} \sum_{(\tx,\mu) \in {\cX}^\text{T}_i \times \cX_i^\text{f}} \Li_2 \left( \frac{\mu}{\tx} \right)
  \right)
  & (q_1 \to 1)
  \\[1.5em]
  \displaystyle
  \exp
  \left(
   \frac{1}{\epsilon_2} \sum_{(x,\mu) \in {\cX}_i \times \cX_i^\text{f}} \Li_2 \left( \frac{\mu}{x} \right)
  \right)
  & (q_2 \to 1)  
 \end{cases}
\end{align}
\end{subequations}
In the asymptotic limit of five dimensional quiver gauge theories, we clearly observe that except for the bifundamental hypermultiplets, for all other contributions, NS$_1$ and NS$_2$ limits are simply related via $\epsilon_2 \leftrightarrow d_i \epsilon_1$ exchange as inherited from the full partition functions.
For the bifundamentals, there are multiple contributions with shifted masses in the limit NS$_2$ limit $q_2 \to 1$, while all the contributions are degenerated in the NS$_1$ limit $q_1 \to 1$, so that it becomes symmetric under the exchange of the quiver gauge nodes $i \leftrightarrow j$.
\paragraph{}
Using these asymptotic expressions, we can rewrite the {non-perturbative/instanton} part of the total partition function in terms of the following expressions: 
\bea
\log \Xi_{\cX^\text{T}}^{\text{NS1}}(\tx) 
 &=& \sum_{i\in  \Gamma_0}\frac{1}{d_i \epsilon_1} \Bigg\{
  \sum_{(\alpha,k) \neq (\alpha',k')}
 \frac{1}{2}\left( \Li_2\left(\frac{\tx_{i,\alpha, k}}{\tx_{i, \alpha', k'}}q_2\right) -  \Li_2\left(\frac{\tx_{i,\alpha, k}}{\tx_{i, \alpha', k'}}\frac{1}{q_2}\right) \right)
 +{2\pi i \tau_i} \sum_{\alpha=1}^{n_i} \sum_{k=1}^\infty \log \tx_{i, \alpha, k} 
\nonumber\\
&+&\sum_{\alpha=1}^{n_i} \sum_{k=1}^\infty \sum_{\tmu \in \cX_i^{\text{af}}}\Li_2 \left(\frac{\tmu}{\tx_{i, \alpha, k}}\right)
 -\sum_{\alpha=1}^{n_i} \sum_{k=1}^\infty \sum_{\mu \in \cX_i^{\text{f}}}\Li_2 \left(q_2\frac{\tx_{i, \alpha, k}}{\mu}\right)\Bigg\}\nonumber\\
&-&\sum_{e\in \Gamma_1}   \sum_{\alpha=1}^{n_i} \sum_{\alpha'=1}^{n_j} \sum_{k=1}^{\infty} \sum_{k'=1}^{\infty} 
 \frac{1}{d_{ij}\epsilon_1}L\left(\frac{1}{\mu_e}\frac{\tx_{i,\alpha, k}}{\tx_{j, \alpha', k'}}; q_2\right)
\eea

\bea
\log \Xi_{\cX}^{\text{NS2}}(x) 
 &=& \sum_{i\in  \Gamma_0}\frac{1}{\epsilon_2} \Bigg\{
 \sum_{(\alpha,k) \neq (\alpha',k')}
 \frac{1}{2}\left( \Li_2\left(\frac{x_{i,\alpha, k}}{x_{i, \alpha', k'}}q_1^{d_i}\right) -  \Li_2\left(\frac{x_{i,\alpha, k}}{x_{i, \alpha', k'}}\frac{1}{q_1^{d_i}}\right)\right) 
 +{2\pi i \tau_i} \sum_{\alpha=1}^{n_i} \sum_{k=1}^\infty \log x_{i, \alpha, k} 
\nonumber\\
&+&\sum_{\alpha=1}^{n_i} \sum_{k=1}^\infty \sum_{\tmu \in \cX_i^{\text{af}}}\Li_2 \left(\frac{\tmu}{x_{i, \alpha, k}}\right)
 -\sum_{\alpha=1}^{n_i} \sum_{k=1}^\infty \sum_{\mu \in \cX_i^{\text{f}}}\Li_2 \left(q_1^{d_i}\frac{x_{i, \alpha, k}}{\mu}\right)\Bigg\}\nonumber\\
&-&\sum_{e\in \Gamma_1}   \sum_{\alpha=1}^{n_i} \sum_{\alpha'=1}^{n_j} \sum_{k=1}^{\infty} \sum_{k'=1}^{\infty} \sum_{r = 0}^{d_i/d_{ij}-1}
 \frac{1}{\epsilon_2}L\left(\frac{1}{\mu_e}\frac{x_{i,\alpha, k}}{x_{j, \alpha', k'}}q_1^{-r d_{ij}}; q_1^{d_j}\right)
\eea

\subsection{$BC_2$ quiver}
\paragraph{}
Considering our prototype example, the $BC_2$ quiver gauge theory.
In this case the bifundamental contribution behave in the NS$_{1,2}$ limits as follows:
\begin{subequations}
\begin{align}
 Z_{1 \to 2}^\text{bf}
 & \to
 \begin{cases}
  \displaystyle
  \exp
  \left(
  - \frac{1}{\epsilon_1}
  \sum_{(x,x') \in \tilde{\cX}_2 \times \tilde{\cX}_1}
  \left(
  L(\mu q_2^{-1} \frac{x}{x'};q_2^2)
  + L(\mu q_2^{-2} \frac{x}{x'};q_2^2)  
  \right)
  \right)
  & (q_1 \to 1) \\
  \displaystyle
  \exp
  \left(
  - \frac{1}{\epsilon_2}
  \sum_{(x,x') \in \tilde{\cX}_2 \times \tilde{\cX}_1}
  \left(
  L(\mu q_1^{-1} \frac{x}{x'};q_1)
  \right)
  \right)
  & (q_2 \to 2)  
 \end{cases} 
 \\[1em]
 Z_{2 \to 1}^\text{bf}
 & \to
 \begin{cases}
  \displaystyle
  \exp
  \left(
  - \frac{1}{\epsilon_1}
  \sum_{(x,x') \in \tilde{\cX}_1 \times \tilde{\cX}_2}
  \left(
  L(\mu^{-1} \frac{x}{x'};q_2)
  \right)
  \right)
  & (q_1 \to 1) \\
  \displaystyle
  \exp
  \left(
  - \frac{1}{\epsilon_2}
  \sum_{(x,x') \in \tilde{\cX}_1 \times \tilde{\cX}_2}
  \left(
  L(\mu^{-1} \frac{x}{x'};q_1)
  \right)
  \right)
  & (q_2 \to 2)  
 \end{cases}
\end{align}
\end{subequations}
In the NS$_1$ limit, it is not symmetric under $1 \leftrightarrow 2$ node exchange due to the duplicated contribution in $Z_{1 \rightarrow 2}^\text{bf}$ with the shifted masses.
On the other hand, in the NS$_2$ limit, we obtain a symmetric situation, which suggests a relation to simply laced quiver.
We will show that it is nothing but the folding of simply laced quiver to obtain a non-simply laced quiver.
\paragraph{}
The dilogarithm function $\Li_2(z)$ 
appearing in the asymptotic limit of five
dimensional vector multiplet, fundamental and antifundamental hyper-multiplet contributions allows us to regard them respectively as the twisted superpotential for the three 
dimensional adjoint, fundamental and antifundamental chiral multiplets compactified on $\bR^2\times S^1$ 
with the shift in twisted masses depending on the remaining deformation parameters $\epsilon_2$ or $d_i \epsilon_1$.
While for the bifundamental hypermultiplet, it descends into different bifundamental chiral multiplet contributions to the twisted superpotential in the compactified theories, depending on which asymptotic limit is taken. It is interesting to note that in the NS$_1$ limit, we obtain multiplet copies of bifundamental chiral multiplets while in the NS$_2$ limit, they degenerate into single copy, 
as the result we can obtain different quantum integrable system in each of these asymptotic limits. We will provide a general discussion about the connection between quantum integrable systems and quiver gauge theories using so-called $qq$-character next.

\section{Doubly quantum Seiberg--Witten curve: $qq$-character}\label{sec:qq-ch}
\paragraph{}
It has been shown that the Seiberg--Witten curve for the quiver gauge theory has an interesting representation theoretical interpretation:
It is nicely described by the fundamental characters associated with simply laced quiver~\cite{Nekrasov:2012xe}, which is also identified with the spectral curve of the associated algebraic classical integrable system. 
Turning on (one of) the $\Omega$-background equivariant deformation parameter, the Seiberg--Witten curve, which is an algebraic curve, is promoted to a difference {{operator}}, called the quantum Seiberg--Witten curve. Again such a quantum geometry can be connected with the representation theory of quiver, where the characters are replaced with the $q$-character of quantum affine algebra associated with the quiver~\cite{Frenkel:1998,Nekrasov:2013xda}. This allows further quantum deformation using the remaining $\Omega$-background parameter, and the corresponding doubly quantum Seiberg--Witten curve is interpreted as the $qq$-characters~\cite{Nekrasov:2015wsu,Kimura:2015rgi}, which can be reduced to the $q$-character in the NS limit, and to the ordinary character in the classical limit $(\epsilon_1,\epsilon_2) \to (0,0)$. 
\paragraph{}
Such a geometric connection between gauge theory and integrable system has been discussed in particular for simply laced quiver, and recently generalized to non-simply laced quiver gauge theory using the fractional quiver~\cite{Kimura:2017hez}.
In the following we discuss how quantum integrable system associated with non-simply laced quiver arises from the quiver gauge theory.

\subsection{Preliminaries to $qq$-character}
\paragraph{}
Under the correspondence between the gauge theory and the integrable system, the gauge invariant observable, namely the chiral ring operator, is identified with the conserved Hamiltonians~\cite{Gorsky:1995zq,Martinec:1995by,Donagi:1995cf}.
From this point of view, the Seiberg--Witten spectral curve, described by the fundamental characters associated with the quiver, plays a role as the generating function of such operators.
After the quantum deformation, the ordinary character is replaced with the $q$-character, which is then identified with the transfer matrix, the generating function of the quantum conserved Hamiltonians in the corresponding quantum integrable system.
In the doubly quantum system, the $qq$-character plays a similar role:
In this case, it is a generating function of the generators for W-algebra associated with the quiver, called the quiver W-algebra.
Since the generators of W-algebra obey nontrivial commutation relations, they are not interpreted as the {{commuting}} Hamiltonians anymore.
However, in general, such non-commuting generators become commuting Hamiltonians, and one obtains quantum integrable system from taking the semi-classical limit of W-algebra, 
corresponding to the NS limit~\cite{Nekrasov:2013xda}.
In the NS limit, the $qq$-character is reduced to the $q$-character of quantum affine algebra associated with quiver.
The $\sT$-operator, obtained as the universal $q$-character, is then identified with the transfer matrix of the corresponding quantum integrable system, obeying the TQ-relation and also the functional relation called the T-system. We will see how such a connection with quantum integrable system is generalized to non-simply laced quiver gauge theory in the following.

\subsubsection{Operator formalism}\label{sec:op_formalism}
\paragraph{}
To construct the $qq$-character from a quiver gauge theory, it is convenient to use the so-called operator formalism:
We consider the deformation of the pre-potential with all the possible chiral ring operators~\cite{Marshakov:2006ii,Nakajima:2003uh}:
\begin{align}
 \msF_\text{UV}
 \ \longrightarrow \
 \msF_\text{UV} + \sum_{i \in \Gamma_0} \sum_{n=1}^\infty t_{i,n} \, \mathcal{O}_{i,n}
 \, ,
\end{align}
where these operators are defined using the adjoint complex scalar in the $i$-th vector multiplet as $\cO_{i,n} = \Tr \Phi_i^n$.
We can compute the deformed partition function now depending on the infinitely many coupling constants $(t_{i,n})_{i \in \Gamma_0, n=1\ldots \infty}$ denoted by $Z(t)$, 
which can be similarly expressed in terms of the summation over the instanton configuration,
\begin{align}
 Z(t) = \sum_{\cX} Z_\cX^\text{tot} Z_\cX^\text{pot}(t), \quad Z(0) = Z
 \label{eq:Z(t)}
\end{align}
where 
\begin{align}
 Z^\text{pot}_\cX(t)
 = \exp \left(
  \sum_{i \in \Gamma_0} \sum_{n=1}^\infty t_{i,n} \, \cO_{i,n}\Big|_{\cX}
 \right)
 \, .
\end{align}
This $t$-dependent part in the deformed partition function is called the {\em potential term} from the analogy with the matrix model,
and plays a role of the chiral ring generating function in the following sense:
\begin{align}
 \Big< \mathcal{O}_{i,n} \Big>
 = \frac{\partial}{\partial t_{i,n}} \log Z(t)\Big|_{t=0}
 \, .
\end{align}
The average here is taken with the Nekrasov partition function associated with the corresponding matter content: 
\begin{align}
 \Big< \mathcal{O} \Big>
 = \frac{1}{{Z}} \sum_{\cX} \mathcal{O}_{\cX} \, Z_\cX^\text{tot}
\end{align}
where $\mathcal{O}_\cX = \left.\mathcal{O}\right|_{\cX}$ denotes the contribution of instanton configuration $\cX$ to expectation value of $\cO$.
Comparing the two expressions above and viewing $Z(t)$ as a function of coupling constants $(t_{i, n})$, the chiral ring operator $(\mathcal{O}_{i,n})$ is promoted to a derivative operator with respect to the conjugate variable $(t_{i,n})$, their commutator generates the Heisenberg algebra $\left[ \partial_{t_{i,n}}, t_{j,n'}\right] = \delta_{ij} \delta_{n,n'}$.
\paragraph{}
We can next generate the Fock space from this Heisenberg algebra,  the $t$-deformed partition function is promoted to an operator acting on this Fock space. Through the operator/state correspondence, we can further map it to a state that we call the $Z$-state:
\begin{align}
 Z(t) \ \longrightarrow \ \ket{Z}
 := Z(t) \ket{1}
 \, ,
\end{align}
where $\ket{1}$ is the vacuum state, annihilated by any derivative operators, $\partial_{t_{i,n}} \ket{1} = 0$, and its dual obeys $\bra{1} t_{i,n} = 0$ for $\forall(i,n)$.
Since the partition function is given as a summation over the instanton configuration \eqref{eq:Z(t)}, the $Z$-state has a similar expression
\begin{align}
 \ket{Z}
 = \sum_{\cX} Z_\cX^\text{tot} Z_\cX^\text{pot}(t) \ket{1}
 =: \sum_{\cX} \ket{Z_\cX} 
\end{align}
where $\cX$ stands for the instanton configuration associated with the fixed point in the instanton moduli space.
Since the dual vacuum $\bra{1}$ plays a role of a projection operator into the undeformed sector $(t=0)$, the undeformed partition function is given by the correlation function:
\begin{align}
 Z(t = 0)
 = \vev{1|Z}
 =
  \sum_{\cX} \vev{1|Z_\cX}
\end{align}
where $\vev{1|Z_\cX} = Z^\text{tot}_\cX$ defined in \eqref{eq:Z_tot}.
See \cite{Kimura:2015rgi} for details.

\subsubsection{Universal $qq$-character}
\paragraph{}
Since the $qq$-character is $(q_1,q_2)$ deformation of the Seiberg--Witten curve, it depends on the Coulomb moduli, the gauge coupling, the (anti)fundamental mass parameters and the Chern--Simons level in 5d.
For example, for $A_1$ quiver gauge theory ($SU(n)$ SQCD with Chern--Simons level $\kappa$ in 5d), the $qq$-character, which is a functional relation, is given by
\begin{align}
 \Big< \sY_{1,x} \Big>
 + \frkq \, x^\kappa \,
 \Big< \sY_{1,q_1^{-1} q_2^{-1} x}^{-1} \Big>
 \times \text{(matter polynomial)}
 = T_{1,n}(x;q_1,q_2)
 \label{eq:qq-ch_A1}
\end{align}
where we introduce the $\sY$-operator, a generating function of the chiral ring operators
\begin{align}
 \sY_{i,x} = \exp \left( - \sum_{s=1}^\infty \frac{x^{-s}}{s} \, \cO_{i,s} \right)
 \, .
\end{align}
Here $T_{1,n}(x;q_1,q_2)$ is a polynomial in $x$, whose degree is given by the gauge group rank $n$, and coefficients depend on the parameters $(q_1,q_2)$.
We denote the exponential of gauge coupling by $\frkq = \exp(2\pi \ci \tau)$, and ``matter polynomial'' is a polynomial factor depending on the (anti)fundamental mass parameters defined later in \eqref{eq:matter_pol}.
In the operator formalism, the $\sY$-operator $\sY_{i,x}$ becomes actually a vertex operator, which shifts the $t$-variable as $t_{i,n} \to t_{i,n} - \frac{x^{-n}}{n}$ because the chiral ring operator $\cO_{i,n}$ is replaced with the derivative $\partial_{t_{i,n}}$ in the operator formalism.
\paragraph{}
In the operator formalism, the $qq$-character is rephrased as follows:
\begin{align}
 \bra{1}
 \left(
 \sY_{1,x} + \sY_{1,q_1^{-1} q_2^{-1} x}^{-1} 
 \right)
 \ket{Z}
 = T_{1,n}(x;q_1,q_2)
\end{align}
where the information about the matter content is encoded within the $Z$-state $\ket{Z}$.
From this point of view, the so-called {\em universal $qq$-character} is obtained as a correlator of a genuine operator lift of the $qq$-character:
\begin{align}
 \sY_{1,} + \sY_{1,q_1^{-1} q_2^{-1} x}^{-1} = \sT_{1,x}
 \, .
\end{align}
Taking the average of this operator relation, we reproduce the $qq$-character~\eqref{eq:qq-ch_A1}, since the average of the $\sT$-operator defined by the universal $qq$-character is reduced to the polynomial $\bra{1} \sT_{1,x} \ket{Z} = T_{1,n}(x;q_1,q_2)$, and the matter polynomial and the Chern--Simons factor are accordingly obtained by taking the average with respect to the corresponding Nekrasov function.
For $A_1$ quiver, the $\sT$-operator has the mode expansion
\begin{align}
 \sT_{1,x} = \sum_{n \in \mathbb{Z}} T_{1,n} \, x^{-n}
\end{align}
where $(T_{1,n})_{n \in \mathbb{Z}}$ is identified with the generator of the $q$-deformed Virasoro algebra~\cite{Shiraishi:1995rp,Frenkel:1996}.

\subsubsection{iWeyl reflection}
\paragraph{}
In general, any representation is generated from the highest weight with the Weyl reflection.
One can apply a similar formalism to systematically construct the universal $qq$-character with the quantum deformation of Weyl reflection, called the ``iWeyl reflection'':
\begin{align}
 \sY_{i,x}
 \ \longrightarrow \
 : \sY_{i,x} \, \sA_{i,q_1^{-1} q_2^{-1} x}^{-1} :
 \quad \text{for} \quad
 i \in \Gamma_0
 \label{eq:iWeyl_ref}
\end{align}
where the normal ordered product is defined such that all the annihilation operators, namely the derivatives with $t$-variables are moved to right, and the creation operators, the $t$-variables themselves, are moved to left.
The $\sY$-operator plays a role of the weight, and we define the root-like operator, called the $\sA$-operator, 
\begin{align}
 \sA_{i,x} =
 \sY_{i,x} \sY_{i,q_1 q_2 x} 
 \left(
 \prod_{e:i \to j} \sY_{j,\mu_e x}
 \prod_{e:j \to i} \sY_{j,\mu_e^{-1} q_1 q_2 x} 
 \right)^{-1}
 \, .
\end{align}
Then we may construct the universal $qq$-character for the representation characterized by the co-weight $w=(w_i)_{i \in \Gamma_0}$:
\begin{align}
 \sT_{w,x}
 = \ :\prod_{i \in \Gamma_0} \prod_{f=1}^{w_i} \sY_{i,x_{i,f}} : + \cdots
 \, .
\end{align}
The lower terms are obtained by the iWeyl reflection \eqref{eq:iWeyl_ref}.
See \cite{Nekrasov:2015wsu} for the geometric construction of the $qq$-character based on the quiver variety.
\paragraph{}
The fundamental $qq$-character is obtained with $\displaystyle w = (0 \ldots \underbrace{1}_{i\text{-th}} \ldots 0)$, namely the highest weight is given by a single $\sY$-operator:
\begin{align}
 \sT_{i,x} = \sY_{i,x} \, + 
 : \sY_{i,x} \, \sA_{i,q_1^{-1} q_2^{-1} x}^{-1} :
 + \cdots
 \, .
 \label{eq:T_exp}
\end{align}
It has been shown in \cite{Kimura:2015rgi} that for quiver $\Gamma$, the $\sT$-operators $(\sT_{i,x})_{i \in \Gamma_0}$ are the generating current of the quiver W-algebra $W_{q_1,q_2}(\Gamma)$, c.f. the energy momentum tensor for the Virasoro algebra.
The expansion \eqref{eq:T_exp} terminates within finite terms if the quiver $\Gamma$ coincides with the Dynkin diagram of finite-type Lie algebra.
Otherwise, the $qq$-character becomes an infinite series.
It has been shown that the $qq$-character $T_i(x) = \bra{1} \sT_{1,x} \ket{Z}$ is a regular pole-free function, which is a polynomial in $x$~\cite{Nekrasov:2015wsu}, and its operator-lift $\sT_{i,x}$ commutes with the screening charges associated with quiver $\Gamma$, which ensures that the $qq$-character is the generating current of the quiver W-algebra~\cite{Kimura:2015rgi}.
We remark that this iWeyl reflection will play a key role on the connection between the SUSY vacuum condition and the Bethe equation discussed in Sec.~\ref{sec:vac}.
\paragraph{}
From the gauge theory perspective, the ``i'' in the iWeyl reflection stands for ``instanton''~\cite{Nekrasov:2015wsu},  as we will shortly explain, the operator $\sA_{i, x}$ which performs the iWeyl reflection precisely corresponds the difference operator which increase the instanton number for a given $\cX_i$ by one, as we will discuss in next section.

\subsection{Non-simply laced quiver: $BC_2$ quiver}
\paragraph{}
It has been recently shown that the $qq$-character and its operator lift, universal $qq$-character can be similarly formulated for non-simply laced quiver gauge theory~\cite{Kimura:2017hez}.
Let us focus on the $BC_2$ quiver as the simplest illustrative example.
In this case, the iWeyl reflection is given by
\begin{align}
 \sY_{1,x}
 \ \longrightarrow \
 \frac{\sY_{2,\mu^{-1} x} \sY_{2,\mu^{-1} q_1^{-1} x}}
      {\sY_{1,q_1^{-2} q_2^{-1} x}}
 \, , \qquad
 \sY_{2,x}
 \ \longrightarrow \
 \frac{\sY_{1,\mu q_1^{-1} q_2^{-1} x}}{\sY_{2,q_1^{-1} q_2^{-1} x}}
\end{align}
where the bifundamental mass is defined $\mu := \mu_{1 \to 2} = \mu_{2 \to 1}^{-1} q_1 q_2$.
The fundamental universal $qq$-characters are respectively generated by these reflections as follows:
\begin{subequations}
\begin{align}
 \sT_{1,x}
 & =
 \sY_{1,x}
 +
 \frac{\sY_{2,\mu^{-1} x} \sY_{2,\mu^{-1} q_1^{-1} x}}
      {\sY_{1,q_1^{-2} q_2^{-1} x}}
 + \msS(q_1)
 \frac{\sY_{2,\mu^{-1}x}}{\sY_{2,\mu^{-1} q_1^{-2} q_2^{-1} x}}
 +
 \frac{\sY_{1,q_1^{-1}q_2^{-1}x}}
      {\sY_{2,\mu^{-1} q_1^{-1} q_2^{-1} x} \sY_{2,\mu^{-1} q_1^{-2} q_2^{-1} x}}
 + \frac{1}{\sY_{1,q_1^{-2} q_2^{-3} x}}
 \\
 \sT_{2,x}
 & =
 \sY_{2,x}
 + \frac{\sY_{1,\mu q_1^{-1} q_2^{-1} x}}{\sY_{2,q_1^{-1} q_2^{-1} x}}
 + \frac{\sY_{2,q_1^{-2} q_2^{-1} x}}{\sY_{1,\mu q_1^{-3} q_2^{-2} x}}
 + \frac{1}{\sY_{2,q_1^{-3} q_2^{-2} x}}
\end{align}
\end{subequations}
where the $\msS$-factor is defined
\begin{align}
 \msS(x) = \frac{(1 - q_1 x)(1 - q_2 x)}{(1 - x)(1 - q_1 q_2 x)}
 \ \stackrel{x \to q_1}{\longrightarrow} \
 \frac{(1 + q_1)(1 - q_1 q_2)}{1 - q_1^2 q_2}
 \, .
 \label{eq:S-fac}
\end{align}
This $\msS$-factor is necessary to ensure the regularity of the $qq$-character, which is obtained as the OPE factor between $\sA$ and $\sY$ operators.
These $\sT$-operators, $\sT_{1,x}$ and $\sT_{2,x}$, correspond to 5-dim vector and 4-dim spinor representations of $BC_2$ quiver, respectively.
We remark that these universal $qq$-characters are identified with the generating currents of the $q$-deformed W$(BC_2)$-algebra~\cite{Frenkel:1997,Bouwknegt:1998da}.
\paragraph{}
Since two $\Omega$-background parameters are not equivalent in non-simply laced quiver gauge theories, we have two possible reductions that we call NS$_1$ $(\epsilon_1 \to 0; q_1 \to 1)$ and NS$_2$ $(\epsilon_2 \to 0; q_2 \to 1)$ limits:
The $\msS$-factor appearing in the $qq$-character actually shows different results
\begin{align}
 \msS(q_1) =  \frac{(1 + q_1)(1 - q_1 q_2)}{1 - q_1^2 q_2}
 \ \longrightarrow \
 \begin{cases}
  2 & (q_1 \to 1) \\
  1 & (q_2 \to 1)
 \end{cases}
\end{align}
Thus we obtain two reductions:\\[.5em]
\underline{NS$_1$ limit}
\begin{subequations}
\begin{align}
 \sT_{1,x}^{\text{NS}_1}
 & =
 \sY_{1,x}
 +
 \frac{\left(\sY_{2,\mu^{-1} x}\right)^2}
      {\sY_{1,q_1^{-1}x}}
 + 2 \, \frac{\sY_{2,\mu^{-1}x}}{\sY_{2,\mu^{-1} q_2^{-1} x}}
 +
 \frac{\sY_{1,2_1^{-1} x}}
      {\left(\sY_{2,\mu^{-1} q_2^{-1}x}\right)^2}
 + \frac{1}{\sY_{1,q_2^{-2} x}}
 \\
 \sT_{2,x}^{\text{NS}_2}
 & =
 \sY_{2,x}
 + \frac{\sY_{1,\mu q_2^{-1} x}}{\sY_{2,q_2^{-1} x}}
 + \frac{\sY_{2,q_2^{-1} x}}{\sY_{1,\mu q_2^{-2} x}}
 + \frac{1}{\sY_{2,q_2^{-2} x}}
\end{align}
\end{subequations}
\underline{NS$_2$ limit}
\begin{subequations}
\begin{align}
 \sT_{1,x}^{\text{NS}_2}
 & =
 \sY_{1,x}
 +
 \frac{\sY_{2,\mu^{-1} x} \sY_{2,\mu^{-1} q_1^{-1} x}}
      {\sY_{1, q_1^{-2} x}}
 + 
 \frac{\sY_{2,\mu^{-1}x}}{\sY_{2,\mu^{-1} q_1^{-2} x}}
 +
 \frac{\sY_{1, q_1^{-1}x}}
      {\sY_{2,\mu^{-1} q_1^{-1} x} \sY_{2,\mu^{-1} q_1^{-2} x}}
 + \frac{1}{\sY_{1, q_1^{-3} x}}
 \\
 \sT_{2,x}^{\text{NS}_1}
 & =
 \sY_{2,x}
 + \frac{\sY_{1,\mu q_1^{-1} x}}{\sY_{2, q_1^{-1} x}}
 + \frac{\sY_{2, q_1^{-2} x}}{\sY_{1,\mu  q_1^{-3} x}}
 + \frac{1}{\sY_{2, q_1^{-3} x}}
\end{align}
\end{subequations}
In taking NS$_1$ and NS$_2$ limits, we obtain different results for $\sT_{1,x}$, while the same results for $\sT_{2,x}$, up to the argument redefinition.
In particular, the operators $\sT_{1,x}^{\text{NS}_2}$ and $\sT_{2,x}^{\text{NS}_{1,2}}$ coincide with the $q$-characters of $BC_2$ quiver.%
\footnote{%
See, for example, \cite{Kuniba:2010ir} for a lot of examples of $q$-characters.
}
Therefore we conclude the correspondence between the quantum integrable system with $BC_2$ symmetry and quiver gauge theory in NS$_2$ limit.
\paragraph{}
On the other hand, the NS$_1$ limit of the operator $\sT_{1,x}$ involves the colliding term $(\sY_{2,x})^2$, which should be interpreted as a limit, $\lim_{x' \to x} \sY_{2,x} \sY_{2,x'}$.
As discussed below, such a collision term is interpreted as a consequence of the folding trick:
The $BC_2$ algebra can be obtained through the folding trick using the outer automorphism of the associated simply laced algebra $AD_3$.
This is true for the classical case, however, for the quantum case, one has to be careful of the quantum parameter dependence.
In this sense, the NS$_1$ and NS$_2$ limits lead to what we shall name a {\em naively folded} and {\em  $\hbar$-folded} integrable system, respectively.


\section{SUSY vacuum and Bethe equation}
\label{sec:vac}
\paragraph{}
In this section, we study the saddle point configurations which dominate the partition functions in the two asymptotic limits, and their truncations.
\subsection{Saddle point equation}
\paragraph{}
In the NS$_{1,2}$ limit, taking $\epsilon_{1,2} \to 0$, we can apply the saddle point approximation to the partition function.
The critical configuration $\cX_*$ is determined by the saddle point equation with respect to the dynamical variable:
\begin{align}
 \exp \left( \epsilon_{m} \frac{\partial}{\log x} \log Z^\text{tot}_{\cX_*} \right) = 1
 \quad \text{with} \quad
 \epsilon_m \to 0
 \quad
 \left( m = 1, 2 \right)
 \label{eq:saddle_pt_eq}
\end{align}
Let us consider this saddle point condition in the operator formalism.
As explained earlier in the previous section, the partition function deformed by the infinitely many time variables $(t_{i,n})_{i \in \Gamma_0, n \in [1,\ldots,\infty]}$ is promoted to the state in the corresponding Fock space, called the $Z$-state.
See Sec.~\ref{sec:op_formalism}.
Then the matrix element of the $\sA$-operator provides
\begin{align}
 \msA_{i,x} := \vev{1|\sA_{i,x}|Z_\cX}
 =
 \frac{Z_\cX^\text{tot}}{Z_{\cX_{(i)}}^\text{tot}}
 \, ,
\end{align}
where $\cX_{(i)}$ denotes the configuration such that another box is added to only the Young diagram $\cX_i$ (or $\cX_i^\text{T}$) of the $i$-th gauge node, or equivalently, one of the variables is shifted by $q_2$ (or $q_1$):
\begin{align}
 \cX_{(i)}
 = \{ x \to q_2 x, x \in \cX_i;\ x \to x, x \in \cX_{j(\neq i)} \}
 \, .
\end{align}
{{We see that $\sA_{i,x}$ plays the role of shift operator which shifts the number of instanton in the configuration $\cX_i$ by one}}.
Thus, in the NS$_{1,2}$ limit, we have
\begin{align}
 \msA_{i,x}
 \ \longrightarrow \
 \begin{cases}
  \displaystyle
  \exp \left( - \epsilon_2 \frac{\partial}{\log x} \log Z^\text{tot}_{\cX} \right)
  & (\epsilon_2 \to 0, x \in \cX_i)
  \\[1em]
  \displaystyle
  \exp \left( - d_i \epsilon_1 \frac{\partial}{\log \tx} \log Z^\text{tot}_{\cX} \right)
  & (\epsilon_1 \to 0, \tx \in \cX_i^\text{T})  
 \end{cases}
\end{align}
The matrix element $\msA_{i,x}$ also has an alternative expression in terms of $\msY$-functions,
\begin{align}
 \msA_{i,x}^{-1}
 = - \frkq_i \,
 \frac{\sP_{i,q_1^{d_i} q_2 x} \widetilde{\sP}_{i,x}}
      {\msY_{i,q_1^{d_i} q_2 x} \msY_{i,x}}
 \prod_{e:i \to j} \prod_{r=0}^{d_{i}/d_{ij}-1}\msY_{j,\mu_e q_1^{r d_{ij}} x}
\end{align}
where we define $\sP_{i,x}$ and $\widetilde{\sP}_{i,x}$ as the fundamental and antifundamental matter polynomials:
\begin{align}
 \sP_{i,x} = \prod_{\mu \in \cX^\text{f}_i}
 \left( 1 - \frac{x}{\mu} \right)
 \, , \qquad
 \widetilde{\sP}_{i,x} = \prod_{\mu \in \cX^\text{af}_i}
 \left( 1 - \frac{\mu}{x} \right)
 \, ,
 \label{eq:matter_pol}
\end{align}
and the $\msY$-function as:
\begin{align}
 \msY_{i,x}
 =
 \bra{1} \sY_{i,x} \ket{Z_\cX}
 =
 \prod_{x' \in \cX} \frac{1 - x'/x}{1 - q_1 x' / x}
 =
 \prod_{x' \in \cX^\text{T}} \frac{1 - x'/x}{1 - q_2^{d_i} x'/x}
 \, .
 \label{eq:Y-func} 
\end{align}
For the latter convenience, we shift the $\msA$-factor (rescaling the $\msY$-function by a rational function)
\begin{align}
 \msA_{i,x}^{-1}
 = - \frkq_i \, 
 \frac{\widetilde{\sP}_{i,x}}{\sP_{i,x}}
 \frac{1}{\msY_{i,q_1^{d_i} q_2 x} \msY_{i,x}}
 \prod_{e:i \to j} \prod_{r=0}^{d_{i}/d_{ij}-1}\msY_{j,\mu_e q_1^{r d_{ij}} x}
 \, .
 \label{eq:saddle_pt_Aop}
\end{align}
Thus the resultant saddle point equations \eqref{eq:saddle_pt_eq} in the NS limits are given as follows:
\begin{subequations}
 \label{eq:saddle_pt_eq_gen}
 \begin{align}
  \text{NS}_1: \quad &
  1 = - \frkq_i \frac{\widetilde{\sP}_{i,x}}{\sP_{i,x}}
  \frac{1}{\msY_{i,q_2 x} \msY_{i,x}}
  \prod_{e:i \to j} \left( \msY_{j,\mu_e x} \right)^{d_i/d_{ij}}
  \quad \text{for} \quad
  x \in \cX_i^\text{T}
  \\
  \text{NS}_2: \quad &
  1 = - \frkq_i \frac{\widetilde{\sP}_{i,x}}{\sP_{i,x}}
  \frac{1}{\msY_{i,q_1^{d_i} x} \msY_{i,x}}
  \prod_{e:i \to j} \prod_{r=0}^{d_i/d_{ij}-1}
  \msY_{j,\mu_e q_1^{r d_{ij}} x} 
  \quad \text{for} \quad
  x \in \cX_i  
 \end{align} 
\end{subequations}
\paragraph{}
Let us consider $A_1$ quiver now as an example.
The saddle point equation is given by
\begin{align}
 1 =
 - \frkq \, \frac{\widetilde{\sP}_{x}}{\sP_{x}}
 \frac{1}{\msY_{i,q_1 x} \msY_{i,x}}
 \quad \text{for} \quad
 x \in \cX
 \quad
 (q_2 \to 1)
 \, .
\end{align}
In this case, the NS$_1$ and NS$_2$ limits are equivalent to each other under exchanging $(\cX,q_2)$ and $(\cX^\text{T},q_1)$.
Moreover, in the NS limit, we can rewrite the $\msY$-function in terms of the ratio of $Q$-function $Q_x$:
\begin{align}
 \msY_x = \frac{Q_{x}}{Q_{q_1^{-1}x}}
  \quad \text{with} \quad
 {Q}_x =
 \prod_{\mu \in \cX^\text{f}}
 \left( q_1 \frac{x}{\mu}; q_1 \right)_\infty^{-1}
 \prod_{x' \in {\cX}}
 \left( 1 - \frac{x'}{x} \right) 
 \, ,
\end{align}
Another equivalent expression is in terms of $\tilde{Q}$-function:
\begin{align}
 \msY_{x} =
 \frac{\tilde{Q}_x}{\tilde{Q}_{q_2^{-1} x}}
 \quad \text{with} \quad
 \tilde{Q}_x =
 \prod_{\mu \in \cX^\text{f}}
 \left( q_2 \frac{x}{\mu}; q_2 \right)_\infty^{-1} 
 \prod_{x' \in \cX^\text{T}}
 \left( 1 - \frac{x'}{x} \right)
 \, ,
\end{align}
we obtain the same result in the limit $q_1 \to 1$ with $\tilde{Q}$-functions. 
We remark it has some extra factor $\prod_{\mu \in \cX^\text{f}} \left( q_{m} x/\mu; q_{m} \right)_\infty^{-1}$ for $m = 1,2$ within the definitions of $Q_x$ and $\tilde{Q}_x$, which do not have any zeros, due to the shift of the $\msA$-factor mentioned earlier.
Finally in terms of the $Q$-functions, the saddle point equation is given by
\begin{align}
 \frac{\sP_{x}}{\widetilde{\sP}_{x}}
 = - \frkq \, \frac{Q_{q_1^{-1} x}}{Q_{q_1 x}}
 \quad \text{for} \quad
 x \in \cX
\end{align}
where $\cX$ is now interpreted as a set of (infinitely many) Bethe roots as the Young diagrams can grow to infinite size.
This is the $A_1$-type Bethe equation:
The gauge coupling $\frkq$ is the twist parameter, the fundamental mass parameters are the inhomogeneous parameters.
\paragraph{}
More generally, from the saddle point equations \eqref{eq:saddle_pt_eq_gen}, we obtain the Bethe equation in terms of $Q$-functions
\begin{subequations} \label{eq:BE_NSlims}
 \begin{align}
  \text{NS}_1: \quad &
  \frac{\sP_{i,x}}{\widetilde{\sP}_{i,x}}
  = - \frkq_i \,
  \frac{\tilde{Q}_{i,q_2^{-1} x}}{\tilde{Q}_{i,q_2 x}}
  \prod_{e:i \to j}
  \left( \frac{\tilde{Q}_{j,\mu_e q_2^{1/2} x}}{\tilde{Q}_{j,\mu_e q_2^{-1/2} x}} \right)^{d_i/d_{ij}}
  \quad \text{for} \quad
  x \in \cX_i^\text{T}
  \label{eq:BE_NS1}\\[1em]
  \text{NS}_2: \quad &
  \frac{\sP_{i,x}}{\widetilde{\sP}_{i,x}}
  = - \frkq_i \, \frac{Q_{i,q_1^{-d_i} x}}{Q_{i,q_1^{d_i} x}}
  \prod_{e:i \to j} \prod_{r=0}^{d_i/d_{ij} - 1}
  \frac{Q_{j,\mu_e q_1^{(d_j-d_i+d_{ij})/2+rd_{ij}} x}}{Q_{j,\mu_e q_1^{-(d_j-d_i+d_{ij})/2-rd_{ij}} x}}
  \quad \text{for} \quad
  x \in \cX_i
  \label{eq:BE_NS2}
 \end{align}
\end{subequations}
where we shift the bifundamental mass $\mu_e \to \mu_e q_1^{1/2} q_2^{(d_j-d_i+d_{ij})/2}$ using the global $U(1)$ symmetries to obtain more symmetric expression.
In particular, if $d_{ij} = d_j$, the Bethe equation in the NS$_2$ limit is simplified as follows:
\begin{align}
  \text{NS}_2: \quad &
  \frac{\sP_{i,x}}{\widetilde{\sP}_{i,x}}
  = - \frkq_i \, \frac{Q_{i,q_1^{-d_i} x}}{Q_{i,q_1^{d_i} x}}
  \prod_{e:i \to j} 
  \frac{Q_{j,\mu_e q_1^{d_i/2} x}}{Q_{j,\mu_e q_1^{-d_i/2} x}}
  \quad \text{for} \quad
  x \in \cX_i 
\end{align}
This is essentially the Bethe equation for the generic (finite-type) Lie algebra $\frkg$~\cite{Reshetikhin:1987bz}
\begin{align}
 \frac{\sP_{i,x}}{\widetilde{\sP}_{i,x}}
 = - \frkq_i \,
 \prod_{j \in \Gamma_0}
 \frac{Q_{j,q^{-b_{ij}/2} x}}{Q_{j,q^{b_{ij}/2} x}}
 \quad \text{for} \quad
 x \in \cX_{i}
 \label{eq:Bethe_eq_gen}
\end{align}
where $(b_{ij})$ is the symmetrized Cartan matrix associated with $\frkg$.
From this point of view, the saddle point equation \eqref{eq:BE_NS2} is interpreted as a generalized version of the Bethe equation associated with the non-simply laced fractional quiver.
On the other hand, the saddle point equation in the NS$_1$ limit \eqref{eq:BE_NS1} involves higher-order poles with the degree $d_i/d_{ij}$.
Such a peculiar behavior is reasonably understood as the consequence of the folding procedure as discussed below.

\subsection{Bethe equation: $BC_2$ case}
\paragraph{}
Let us apply this analysis to $BC_2$ as a prototype example of non-simply laced quiver gauge theories.
Since we have two nodes in this case, we have two $\msA$-factors \eqref{eq:saddle_pt_Aop}:
\begin{align}
 \msA_{1,x}^{-1} = - \frkq_1 \,
 \frac{\widetilde{\sP}_{1,x}}{\sP_{1,x}}
 \frac{\msY_{2, q_2^{1/2} x} \msY_{2, q_1 q_2^{1/2} x}}
      {\msY_{1,x} \msY_{1,q_1^2 q_2 x}}
 \, , \qquad
 \msA_{2,x}^{-1} = - \frkq_2 \,
 \frac{\widetilde{\sP}_{2,x}}{\sP_{2,x}}
 \frac{\msY_{1, q_1 q_2^{1/2} x}}
      {\msY_{2, x} \msY_{2,q_1 q_2 x}}
 \, .
\end{align}
where we shift the bifundamental mass parameters using the gauge transform to simplify the expression.
In this case the $\msY$-functions are written in terms of $Q$-functions as follows:
\begin{align}
 \msY_{1,x}
 = \frac{Q_{1,x}}{Q_{1,q_1^{-2}x}}
 = \frac{\tilde{Q}_{1,x}}{\tilde{Q}_{1,q_2^{-1}x}}
 \, , \qquad
 \msY_{2,x}
 = \frac{Q_{2,x}}{Q_{2,q_1^{-1}x}}
 = \frac{\tilde{Q}_{2,x}}{\tilde{Q}_{2,q_2^{-1}x}}
 \, .
\end{align}
Thus the sets of Bethe equations are obtained from taking the NS$_1$ and NS$_2$ limit respectively on each of the gauge nodes.
For the second gauge node, we have essentially symmetric sets of saddle equations from each limit:
\begin{subequations}
\begin{align}
 \text{NS}_1: \quad &
 \frac{\sP_{2,x}}{\widetilde{\sP}_{2,x}}
 = - \frkq_2 \,
 \frac{\tilde{Q}_{2,q_2^{-1} x}}{\tilde{Q}_{2,q_2 x}}
 \frac{\tilde{Q}_{1,q_2^{1/2}x}}{\tilde{Q}_{1,q_2^{-1/2}x}}
 & \text{for} \quad
 x \in \cX^\text{T}_2\,,
 \label{eq:BE2_BC2_NS1}
 \\
 \text{NS}_2: \quad &
 \frac{\sP_{2,x}}{\widetilde{\sP}_{2,x}}
 = - \frkq_2 \,
 \frac{Q_{2,q_1^{-1} x}}{Q_{2,q_1 x}}
 \frac{Q_{1,q_1 x}}{Q_{1,q_1^{-1}x}} 
 & \text{for} \quad
 x \in \cX_2
 \, .
 \label{eq:BE2_BC2_NS2}
\end{align}
\end{subequations}
While for the first gauge node, we obtain different sets of saddle point equations from each limit:
\begin{subequations} 
 \begin{align}
  \text{NS}_1: \quad &
 \frac{\sP_{1,x}}{\widetilde{\sP}_{1,x}}
 = - \frkq_1 \,
  \frac{\tilde{Q}_{1,q_2^{-1} x}}{\tilde{Q}_{1,q_2 x}}
  \left( \frac{\tilde{Q}_{2,q_2^{1/2} x}}{\tilde{Q}_{2,q_2^{-1/2} x}} \right)^2
  & \text{for} \quad
  x \in \cX^\text{T}_1\,,
  \label{eq:BE1_BC2_NS1}\\
  \text{NS}_2: \quad &
 \frac{\sP_{1,x}}{\widetilde{\sP}_{1,x}}
 = - \frkq_1 \,
 \frac{{Q}_{1,q_1^{-2} x}}{{Q}_{1,q_1^2 x}}
 \frac{{Q}_{2,q_1 x}}{{Q}_{2, q_1^{-1} x}}
  & \text{for} \quad
  x \in {\cX}_1\,.
  \label{eq:BE1_BC2_NS2}
 \end{align}
\end{subequations}
Since for $BC_2$ quiver the Cartan matrix and its symmetrization are given as
\begin{align}
 c =
 \begin{pmatrix}
  2 & -1 \\ -2 & 2
 \end{pmatrix}
 \, , \qquad
 b =
 \begin{pmatrix}
  2 & \\ & 1
 \end{pmatrix}
 \begin{pmatrix}
  2 & -1 \\ -2 & 2
 \end{pmatrix} 
 =
 \begin{pmatrix}
  4 & -2 \\ -2 & 2 
 \end{pmatrix}
 \, ,
\end{align}
one can see that in NS$_2$ limit, the saddle point equations \eqref{eq:BE2_BC2_NS2} and \eqref{eq:BE1_BC2_NS2} coincide with the Bethe equations of $BC_2$-type from the generic formula~\eqref{eq:Bethe_eq_gen}.

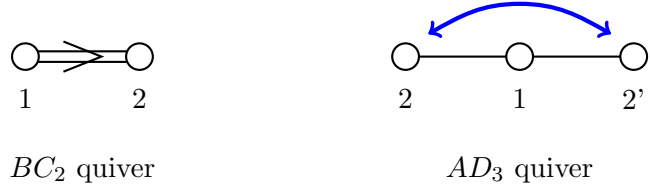
\begin{figure}[t]
\begin{center}
 \begin{tikzpicture}[thick]

  \draw (0,.07) -- ++(1.5,0);
  \draw (0,-.07) -- ++(1.5,0);

  \draw (.5,.2) -- (1,0) -- (.5,-.2);
  
  \filldraw [fill=white] (0,0) circle [radius = 5pt] node [below=.75em] {1};
  \filldraw [fill=white] (1.5,0) circle [radius = 5pt] node [below=.75em] {2};
  \node at (.75,-1.5) {$BC_2$ quiver};

   \begin{scope}[shift={(5,0)}]

    \draw (0,0) -- ++(3,0);
    
    \filldraw [fill=white] (0,0) circle [radius = 5pt] node [below=.75em] {2};
    \filldraw [fill=white] (1.5,0) circle [radius = 5pt] node [below=.75em] {1};
    \filldraw [fill=white] (3,0) circle [radius = 5pt] node [below=.75em] {2'};

    \draw [<->,blue,ultra thick] (.3,.3) to [out=30,in=left] (1.5,.7) to [out=right,in=150] (2.7,.3);

    \node at (1.5,-1.5) {$AD_3$ quiver};
    
   \end{scope}
  
 \end{tikzpicture}
\end{center}
 \caption{From $AD_3$ to $BC_2$ by folding (identifying) the nodes $i=2$ and 2'. The direction of the doubled arrow is from long to short roots.}
 \label{fig:folding}
\end{figure}
\paragraph{}
In the NS$_1$ limit, on the other hand, we obtain a peculiar degenerated factor $\left(Q_{2,x}\right)^2$ in \eqref{eq:BE1_BC2_NS1}, which does not appear in the Bethe equation for any Lie algebra $\frkg$~\eqref{eq:Bethe_eq_gen}.%
\footnote{%
We remark that such a degenerated factor was also found in a analysis of 3d non-simply laced quiver gauge theory~\cite{Dey:2016qqp}.
See also \cite{Ito:2016qzt}.
}
In fact we have an interpretation of such a factor as a consequence of folding $A_3$/$D_3$ quiver, say $AD_3$ quiver for short.
$BC_2$ quiver is obtained from $AD_3$ quiver by folding nodes 2 and 2' as shown in Fig.~\ref{fig:folding}.
Under this identification, the Bethe equation for the node $i=1$ becomes
\begin{align}
 \frac{\sP_{1,x}}{\widetilde{\sP}_{1,x}}
 = - \frkq_1 \,
 \frac{{Q}_{1,q_2^{-1} x}}{{Q}_{1,q_2 x}}
 \frac{{Q}_{2,q_2^{1/2} x}}{{Q}_{2,q_2^{-1/2} x}}
 \frac{{Q}_{2',q_2^{1/2} x}}{{Q}_{2',q_2^{-1/2} x}} 
 \ \longrightarrow \
 - \frkq_1 \,
 \frac{{Q}_{1,q_2^{-1} x}}{{Q}_{1,q_2 x}}
 \left( \frac{{Q}_{2,q_2^{1/2} x}}{{Q}_{2,q_2^{-1/2} x}} \right)^2
 \, ,
\end{align}
where the arrow indicates the folding of the quiver nodes and final expression is consistent with \eqref{eq:BE1_BC2_NS1}.

\subsection{Root of Higgs branch and truncation of Bethe roots}
\paragraph{}
In previous sections, we obtained the Bethe equation from the saddle point equations in the NS$_{1,2}$ limits, which can be written in terms of the $Q$-functions.
The $Q$-functions obtained from the gauge theory have infinitely many zeros at $x \in \cX$, whose values are interpreted as the Bethe roots. 
These Bethe roots are organized into Young diagrams which can generally contain infinite number of rows or columns, and total number of boxes of all Young diagrams is the gauge instanton number.
For simply laced quiver gauge theories, it is known that we can truncate the rows or columns to finite numbers by imposing quantization condition on the root of Higgs branch~\cite{Dorey:2011pa,Chen:2011sj}, and this implies inserting the co-dimension two topological defects, charged under the combined color (gauge) and flavor symmetry groups, into these theories. 
In particular the number of the rows in the truncated Young diagrams can be interpreted as the topological numbers associated with these defects, 
and the on-shell twisted superpotentials for the defect world volume theories can be identified with the free energy of the underlying quantum integrable systems giving the Bethe ansatz equations.
\paragraph{}
Here we show that the similar quantization procedure is applicable to the non-simply laced quiver theory, 
however the interesting subtlety here is that two distinct types of topological defects can arise in two orthogonal co-dimension two planes, and we will identify their corresponding world volume theories in turns.
To illustrate how the quantization condition leads to truncation, let us consider the fundamental hypermultiplet contribution to the 5d partition function given in \eqref{eq:Zf_5d}%
\footnote{%
The discussion for 6d cases is completely parallel.
}
\begin{align}
 Z_i^\text{f}
 =
 \prod_{(\tx,\mu) \in \cX_i^{\text{T}} \times \cX_i^\text{f}}
 \left(\frac{\tx}{\mu};q_1^{d_i}\right)_\infty^{-1}
 \, ,
\end{align}
where we shift the mass parameter $\mu \in \cX^\text{f}_i$ to absorb the additional $q_1^{d_i}q_2$ factor for simplicity.
Applying the analytic continuation
\begin{align}
 (z;q)_\infty
 = (z q^{-1} ; q^{-1})_\infty^{-1}
 \, ,
\end{align}
and explicitly writing out the components which range over the Young diagrams contained in $\cX^{\text{T}}$, we obtain the expression with $n^\text{f}_i$ fundamental matters:
\begin{align}
 Z_i^\text{f}
 & =
 \prod_{\alpha=1}^{n_i} \prod_{f=1}^{n_i^\text{f}} \prod_{k=1}^\infty 
 \left(
 \frac{q_1^{d_i (\lambda_{i,\alpha,k}^{\rm T}-1)} q_2^{k-1} \nu_{i,\alpha}}{\mu_{i,f}};q_1^{-d_i}
 \right)_\infty
 \, .
\end{align}
We now tune the Coulomb moduli parameter to impose the quantization condition at the root of Higgs branch in the NS$_1$ limit~\cite{Dorey:2011pa,Chen:2011sj, Chen:2012we}:
\begin{align}
 \mu_{i,f} = q_2^{\tilde{\rn}_{i,\alpha}} \nu_{i,\alpha}
 \label{eq:Higgs}
\end{align}
which is equivalent to setting $m_{i,f} = a_{i,\alpha} + \tilde{\rn}_{i,\alpha} \epsilon_2$ with the additional constant shifts.
We consider the following factor in the fundamental hypermultiplet contribution:
\begin{align}
 \prod_{k=1}^\infty
 \left(
 q_1^{d_i \lambda_{i, \alpha,k}^{\rm T}} q_2^{k-1-\rn_{i,\alpha}}; q_1^{-d_i}
 \right)_\infty
 =
 \prod_{k=1}^\infty \prod_{r=1}^\infty
 \left(
 1 -  q_1^{d_i (\lambda_{i, \alpha,k}^{\rm T} - r)} q_2^{k-\tilde{\rn}_{i,\alpha}-1}
 \right)
 \, ,
\end{align}
this infinite product actually vanishes when:
\begin{align}
 Z_i^\text{f} = 0
 \quad \text{at} \quad
 k = \tilde{\rn}_{i,\alpha} + 1
 \quad \text{if} \quad
 \lambda_{i,\alpha,k}^{\rm T} \ge 1
 \, .
\end{align}
Therefore, in order to have non-vanishing contribution to the partition function, we have to assign the following truncation condition on the transposed Young diagrams:
\begin{align}
 \lambda_{i,\alpha,k}^{\rm T} = 0
 \quad \text{for} \quad
 k \ge \tilde{\rn}_{i,\alpha} + 1
 \, ,
 \label{eq:partition_truncation}
\end{align}
which is equivalent to the condition for the Young diagrams
\begin{align}
 {\lambda}_{i,\alpha,1} \le \tilde{\rn}_{i,\alpha}
 \, .
\end{align}
This means that the partition $(\lambda_{i,\alpha,k})$, or equivalently the $\tx$-variable $(\tx_{i,\alpha,k})$, becomes non-dynamical for $k \ge \tilde{\rn}_{i,\alpha} + 1$.
Under the condition \eqref{eq:partition_truncation}, the infinite product of the $\msY$-function \eqref{eq:Y-func} is truncated as follows:
\begin{align}
 \msY_{i,x}
 =
 \prod_{\alpha=1}^{n_i}
 \left[
 \left(1 - \frac{\mu_{i,\alpha}}{x} \right) 
 \prod_{k=1}^{\tilde{\rn}_{i,\alpha}}
 \frac{1 - q_2^{d_i \lambda_{i,\alpha,k}} q_1^{k-1} \nu_{i,\alpha}/x}
      {1 - q_2^{d_i \lambda_{i,\alpha,k}} q_1^{k} \nu_{i,\alpha}/x}
 \right]
 \, .
\end{align}
Thus the number of zeros of $Q$-function constructed from this $\msY$-function becomes finite.
We remark that the factor $(1 - \mu_{i,\alpha}/x)$ does not affect the zeros of $Q$-functions.
We can obtain a similar result starting with another equivalent expression for the fundamental hypermultiplet contribution
\begin{align}
 Z_i^\text{f}
 =
 \prod_{(x,\mu) \in \cX_i \times \cX_i^\text{f}}
 \left(\frac{x}{\mu};q_2\right)_\infty^{-1}
 & =
 \prod_{\alpha=1}^{n_i} \prod_{f=1}^{n_i^\text{f}} \prod_{k=1}^\infty 
 \left(
 \frac{q_2^{\lambda_{i,\alpha,k}-1} q_1^{d_i(k-1)} \nu_{i,\alpha}}{\mu_{i,f}};q_2^{-1}
 \right)_\infty 
 \, .
\end{align}
In this case, imposing the quantization condition at the root of Higgs branch for NS$_2$ limit:
\begin{align}
 \mu_{i,f} = q_1^{d_i {\rn}_{i,\alpha}} \nu_{i,\alpha}\label{eq:Higgs2}
 \, ,
\end{align}
the Young diagrams are truncated as follows:
\begin{align}
 \lambda_{i,\alpha,k}= 0
 \quad \text{for} \quad
 k \ge {\rn}_{i,\alpha} + 1
 \quad \iff \quad
 \lambda_{i,\alpha,1}^{\text{T}} \le {\rn}_{i,\alpha}
 \, .
\end{align}
Thus the resultant $Q$-function is again truncated to yield the following resultant $\msY$-function:
\begin{align}
 \msY_{i,x}
 =
 \prod_{\alpha=1}^{n_i}
 \left[
 \left(1 - \frac{\mu_{i,\alpha}}{x} \right) 
 \prod_{k=1}^{{\rn}_{i,\alpha}}
 \frac{1 - q_1^{\lambda_{i,\alpha,k}} q_2^{d_i(k-1)} \nu_{i,\alpha}/x}
      {1 - q_1^{\lambda_{i,\alpha,k}} q_2^{d_i k} \nu_{i,\alpha}/x}
 \right]
 \, ,
\end{align}
which is an equivalent representation.
\paragraph{}
If we now impose these truncation conditions to the saddle point equations arising respectively from the NS$_1$ and NS$_2$ limits, we obtain the finite version of the Bethe equations \eqref{eq:BE_NSlims}:
\begin{subequations} \label{eq:BE_truncated}
 \begin{align}
  \text{NS}_1: \quad &
  \frac{\mathbf{P}_{i,\tx}}{\widetilde{\mathbf{P}}_{i,\tx}}
  = - \hat\frkq_i \,
  \frac{\tilde{\mathbf{Q}}_{i,q_2^{-1} \tx}}{\tilde{\mathbf{Q}}_{i,q_2 \tx}}
  \prod_{e:i \to j}
  \left( \frac{\tilde{\mathbf{Q}}_{j,q_2^{1/2} \tx}}{\tilde{\mathbf{Q}}_{j,q_2^{-1/2} \tx}} \right)^{d_i/d_{ij}}
  \quad \text{for} \quad
  \tx \in \cX_i^\text{T}
  \label{eq:BE_truncated_NS1}
  \\[1em]
  \text{NS}_2: \quad &
  \frac{\mathbf{P}_{i,x}}{\widetilde{\mathbf{P}}_{i,x}}
  = - \hat\frkq_i \, \frac{\mathbf{Q}_{i,q_1^{-d_i} x}}{\mathbf{Q}_{i,q_1^{d_i} x}}
  \prod_{e:i \to j} \prod_{r=0}^{d_i/d_{ij} - 1}
  \frac{\mathbf{Q}_{j,q_1^{(d_j-d_i+d_{ij})/2+rd_{ij}} x}}{\mathbf{Q}_{j,q_1^{-(d_j-d_i+d_{ij})/2-rd_{ij}} x}}
  \quad \text{for} \quad
  x \in \cX_i
  \label{eq:BE_truncated_NS2}
 \end{align}
\end{subequations}
where ${\frkq}_i$ has been shifted to $\hat{\frkq}_i$ to absorb various constants in re-expressing the saddle point equation in terms of following functions:
\begin{align}
 \mathbf{Q}_{i,x}
 & =
 \prod_{\alpha = 1}^{n_i} \prod_{k = 1}^{\rn_{i,\alpha}}
 \sinh \left( \frac{\log x - \log x_{i,\alpha,k}}{2} \right)
 \, , \ 
 \tilde{\mathbf{Q}}_{i,x}
 =
 \prod_{\alpha = 1}^{n_i} \prod_{k = 1}^{\trn_{i,\alpha}}
 \sinh \left( \frac{\log x - \log \tilde{x}_{i,\alpha,k}}{2} \right)
 \, ,\\
 \mathbf{P}_{i,x}
 & = \prod_{f = 1}^{n_i^\text{f}} \sinh \left( \frac{\log x - \log \mu_{i,f}}{2} \right)
 \, , \qquad
 \widetilde{\mathbf{P}}_{i,x}
 = \prod_{f = 1}^{n_i^\text{af}} \sinh \left( \frac{\log x - \log \tilde{\mu}_{i,f}}{2} \right)
 \, .
\end{align}
We remark that the bifundamental mass is removed using the gauge symmetry for simplicity.
Furthermore, if $d_{ij} = d_j$, the saddle point equation in the NS$_2$ limit is reduced as
\begin{align}
  \text{NS}_2: \quad &
  \frac{\mathbf{P}_{i,x}}{\widetilde{\mathbf{P}}_{i,x}}
  = - \hat\frkq_i \, \frac{\mathbf{Q}_{i,q_1^{-d_i} x}}{\mathbf{Q}_{i,q_1^{d_i} x}}
  \prod_{e:i \to j} 
  \frac{\mathbf{Q}_{j,q_1^{d_i/2} x}}{\mathbf{Q}_{j,q_1^{-d_i/2} x}}
  \quad \text{for} \quad
  x \in \cX_i 
\end{align}
which reproduces the Bethe equation associated with generic Lie algebra~\eqref{eq:Bethe_eq_gen}.
Again the saddle point equation \eqref{eq:BE_truncated_NS2} is a generalized version of the Bethe equation associated with the non-simply laced fractional quiver.
\paragraph{}
Here we also comment on how the same sets of Bethe ansatz equations \eqref{eq:BE_truncated}
can also arise from the saddle point equations for the twisted superpotentials of the appropriate $\cN=2$ supersymmetric gauge theories compactified on $\mathbb{R}^2 \times S^1$,
this is in the same vein as the 3d/5d correspondence considered in \cite{Chen:2012we, Chen:2013dda}:
While in our current situation, the corresponding D-brane construction, hence definite interpretation of them as the world volume theories of co-dimension two defects are currently lacking, we nevertheless write down these compactified three dimensional theories for possible references.
\paragraph{}
For NS$_1$ limit \eqref{eq:BE_truncated_NS1}, the corresponding three dimensional $\cN=2$ theory has gauge group $U(K_i)$ with $n_i^{\rm f}$ fundamental chiral multiplets with twisted masses $\log\mu_{i, f}$, $f = 1, \dots, n_i^{\rm f}$; $n_i^{\rm af}$ antifundamental chiral multiplets with twisted masses $\log\tilde{\mu}_{i, f}$, $f = 1, \dots, n_i^{\rm af}$, an adjoint chiral multiplet of twisted mass $\log q_2 = \epsilon_2$, 
finally plus $\frac{d_i}{d_{ij}}$ copies of bifundamental chiral multiplets connecting each pair of gauge group $U(K_i)$ and $U(K_j)$. 
The rank of gauge groups need to satisfy the conditions $\sum_{\alpha=1}^{n_i}\rn_{i, \alpha} =K_i$, 
and we need to impose the similar quantization condition on the vev of the adjoint scalar in each $U(K_i)$ vector multiplet as in \eqref{eq:Higgs}.
Finally the FI parameter for the $U(K_i)$ gauge group is identified up to an unimportant numerical factor with the 5d holomorphic gauge coupling.
\paragraph{}
For NS$_2$ limit \eqref{eq:BE_truncated_NS2}, the corresponding three dimensional $\cN=2$ theory has gauge group $U(K_i)$ with $n_i^{\rm f}$ fundamental chiral multiplets with twisted masses $\log\mu_{i, f}$, $f = 1, \dots, n_i^{\rm f}$; $n_i^{\rm af}$ antifundamental chiral multiplets with twisted masses $\log\tilde{\mu}_{i, f}$, $f = 1, \dots, n_i^{\rm af}$, an adjoint chiral multiplet of twisted mass $d_i\log q_1 = d_i\epsilon_1$, finally for each pair of gauge group $U(K_i)$ and $U(K_j)$, we have $\frac{d_i}{d_{ij}}$ bifundamental chiral multiplets, with varying twisted masses $m_{ij}^{r} = \frac{(d_j-d_i+d_{ij}(r+2))}{2}$, $r = 0, 1, \dots, \frac{d_i}{d_{ij}}-1$.
The rank of gauge groups need to satisfy the conditions $\sum_{\alpha=1}^{n_i}\rn_{i, \alpha} =K_i$, 
and we need to impose the similar quantization condition on the vev of the adjoint scalar in each $U(K_i)$ vector multiplet as in \eqref{eq:Higgs2}.
Finally the FI parameter for the $U(K_i)$ gauge group is again identified up to an unimportant numerical factor with the 5d holomorphic gauge coupling.

\subsection*{Acknowledgements}

We are grateful to P.~Koroteev and V.~Pestun for useful discussions.
The work of HYC was supported in part by in part by Ministry of Science and Technology through the grant 104-2112-M-002 -004-MY and HYC also thanks Keio University for the hospitality when this work was being completed.
The work of TK was supported in part by Keio Gijuku Academic Development Funds, JSPS Grant-in-Aid for Scientific Research (No.~JP17K18090), the MEXT-Supported Program for the Strategic Research Foundation at Private Universities ``Topological Science'' (No. S1511006), JSPS Grant-in-Aid for Scientific Research on Innovative Areas ``Topological Materials Science'' (No.~JP15H05855), and ``Discrete Geometric Analysis for Materials Design'' (No.~JP17H06462).

\appendix
\section{Six dimensional non-simply laced quiver theories}
\label{sec:6d}

\subsection{Partition function}

We provide the analysis for 6d $\mathcal{N} = (1,0)$ theory defined on $\mathbb{R}^4 \times T^2$.
We can apply essentially the same approach as 5d theory on $\mathbb{R}^4 \times S^1$ to obtain the partition function just by replacing the index shown in~\eqref{eq:index}:
\begin{subequations}
\begin{align}
 Z_i^\text{vec} 
  & =
 \prod_{(x,x') \in \cX_i \times \cX_i}
 \Gamma\left(q_1^{d_i} q_2 \frac{x}{x'}; q_2,p\right)^{-1}
 \Gamma\left(q_2 \frac{x}{x'}; q_2, p\right)\,,
 \\
 & =
 \prod_{(\tx,\tx') \in \cX^\text{T}_i \times \cX^\text{T}_i}
 \Gamma\left(q_1^{d_i} q_2 \frac{\tx}{\tx'}; q_1^{d_i},p\right)^{-1}
 \Gamma\left(q_1^{d_i} \frac{\tx}{\tx'}; q_1^{d_i},p\right)
 \, , 
\end{align}
\begin{align} 
 Z_{e:i \to j}^\text{bf}
 & =
 \prod_{(x,x') \in \cX_i \times \cX_j}
 \prod_{r=0}^{d_j/d_{ij}-1}
 \Gamma\left(\mu_e^{-1} q_1^{d_i-rd_{ij}} q_2 \frac{x}{x'}; q_2,p\right)
 \Gamma\left(\mu_e^{-1} q_1^{-rd_{ij}} q_2 \frac{x}{x'}; q_2,p\right)^{-1}\,,
 \\
 & =
 \prod_{(\tx,\tx') \in \cX^\text{T}_i \times \cX^\text{T}_j}
 \prod_{r=0}^{d_j/d_{ij}-1}
 \Gamma\left(\mu_e^{-1} q_1^{d_j-rd_{ij}} q_2 \frac{\tx}{\tx'}; q_1^{d_j},p\right)
 \Gamma\left(\mu_e^{-1} q_1^{d_j-rd_{ij}} \frac{\tx}{\tx'}; q_1^{d_j},p\right)^{-1}
 \, ,
\end{align}
\begin{align} 
 Z_i^\text{f}
 & =
 \prod_{(x,\mu) \in \cX_i \times \cX_i^\text{f}}
 \Gamma\left( q_1^{d_i} q_2 \frac{x}{\mu}; q_2,p \right)
 =
 \prod_{(\tx,\mu) \in \cX^\text{T}_i \times \cX_i^\text{f}}
 \Gamma\left( q_1^{d_i} q_2 \frac{\tx}{\mu}; q_1^{d_i},p \right)
 \, , \\[.5em]
 Z_i^\text{af}
 & =
 \prod_{(\mu,x) \in \cX_i^\text{af} \times \cX_i}
 \Gamma\left( q_2 \frac{\mu}{x}; q_2,p \right)^{-1}
 =
 \prod_{(\mu,\tx) \in \cX_i^\text{af} \times \cX^\text{T}_i}
 \Gamma\left( q_1^{d_i} \frac{\mu}{\tx}; q_1^{d_i},p \right)^{-1}
 \, ,
 \end{align}
\end{subequations}
where we define the elliptic gamma function
\begin{align}
 \Gamma(z;p,q)
 = \prod_{n,m \ge 0} \frac{1 - z^{-1} p^{n+1} q^{n+1}}{1 - z p^n q^n}
 \ \longrightarrow \
 \begin{cases}
  (z;q)_\infty^{-1} & (p \to 0) \\
  (z;p)_\infty^{-1} & (q \to 0)   
 \end{cases}
 \, .
\end{align}
Here we observe that the asymmetry under the exchange of $(q_1^{d_i}, \cX_i^{\text{T}}) \leftrightarrow (q_2, {\cX}_i)$ only occurs in the bifundamental hypermultiplet contributions.

\subsection{Asymptotic behavior}
\paragraph{}
We begin with the following expansion of the elliptic gamma function:
\begin{align}
 \Gamma(z;p,q)
 =
 \prod_{n,m \ge 0}
 \frac{1 - z^{-1} p^{n+1} q^{n+1}}{1 - z p^n q^n}
 =
 \exp
 \left(
  \sum_{m \neq 0} \frac{z^m}{m(1 - p^m)(1 - q^m)}
 \right)
 \, .
\end{align}
Putting $q = e^\epsilon$, the asymptotic behavior in the limit $\epsilon \to 0$ is given by
\begin{align}
 \Gamma(z;p,q)
 & =
 \exp
 \left(
  - \frac{1}{\epsilon} \sum_{m \neq 0} \frac{z^m}{m^2(1 - p^m)}
  + O(\epsilon^0)
 \right)
 \nonumber \\
 & =
 \exp
 \left(
  - \frac{1}{\epsilon} \Li_2(z;p)
  + O(\epsilon^0)
 \right)
 \, ,
\end{align}
where we define an elliptic analogue of the polylogarithm
\begin{align}
 \Li_k(z;p) = \sum_{m \neq 0} \frac{z^m}{m^k(1-p^m)}
 \, .
\end{align}
This is reduced to the ordinary polylogarithm in the limit $p \to 0$ ($\operatorname{Im} \tau \to \infty$), and obeys the similar descendant relation
\begin{align}
 \frac{d}{d \log z} \Li_k(z;p) = \Li_{k-1}(z;p)
 \, .
\end{align}
In particular, we have the following relation with the theta function \eqref{eq:th_fn}:
\begin{align}
 \Li_{1}(z;p)
 = \sum_{m \neq 0} \frac{z^m}{m(1-p^m)}
 = - \log \theta(z;p)
 \, .
\end{align}
Then the elliptic gamma function ratio used in the 6d gauge theory partition function has the asymptotic behavior
\begin{subequations} 
 \begin{align}
 \frac{\Gamma(q_1 q_2^{d_i }z; p, q_2^{d_i})}{\Gamma(q_2^{d_i} z; p, q_2^{d_i})}
 & \stackrel{\epsilon_2 \to 0}{\longrightarrow} \
 \exp
 \left(
  - \frac{1}{d_i \epsilon_2} \left( \Li_2 (q_1 z;p) - \Li_2(z;p) \right)
 \right)
 =
 \exp
 \left(
  - \frac{1}{d_i \epsilon_2} L(z;q_1;p)
 \right)
  \\
 \frac{\Gamma(q_1 q_2^{d_i}z; p, q_1)}{\Gamma(q_1 z; p, q_1)}
 & \stackrel{\epsilon_1 \to 0}{\longrightarrow} \
 \exp
 \left(
  - \frac{1}{\epsilon_1} \left( \Li_2 (q_2^{d_i} z;p) - \Li_2(z;p) \right)
 \right)
 =
 \exp
 \left(
  - \frac{1}{\epsilon_1} L(z;q_2^{d_i};p)
 \right)  
 \end{align}
\end{subequations} 
with the elliptic $L$-function defined
\begin{align}
 L(z;q;p)
 =
 \Li_2 (q z;p) - \Li_2(z;p)
 \, .
\end{align}
The leading contribution of the partition function is obtained from the asymptotic expansion shown above:
\begin{subequations}
\begin{align}
 Z_i^\text{vec}
 & \longrightarrow \
 \begin{cases}
  \displaystyle
  \exp
  \left(
  \frac{1}{\epsilon_1}
  \sum_{(x,x') \in \tilde\cX_i \times \tilde\cX_i}
  L \left(\frac{x}{x'};q_2^{d_i};p\right)
  \right)
  & (q_1 \to 1)
  \\[1.5em]
  \displaystyle
  \exp
  \left(
  \frac{1}{d_i \epsilon_2}
  \sum_{(x,x') \in \cX_i \times \cX_i}
  L \left(\frac{x}{x'};q_1;p\right)
  \right)
  & (q_2 \to 1)
 \end{cases}
\end{align}
\begin{align}
 Z_{e:i \to j}^\text{bf}
 & \longrightarrow \
 \begin{cases}
  \displaystyle
  \exp
  \left(
  - \frac{1}{\epsilon_1}
  \sum_{(x,x') \in \tilde\cX_i \times \tilde\cX_j}
  \sum_{r = 0}^{d_j/d_{ij} - 1}
  L \left( \mu_e^{-1} q_2^{-rd_{ij}} \frac{x}{x'};q_2^{d_j};p \right)
  \right)
  & (q_1 \to 1)
  \\[1.5em]
  \displaystyle
  \exp
  \left(
  - \frac{1}{d_{ij} \epsilon_2}
  \sum_{(x,x') \in \cX_i \times \cX_j}
  L \left( \mu_e^{-1} \frac{x}{x'};q_1;p \right)
  \right)
  & (q_2 \to 1)
 \end{cases}
\end{align}
\begin{align}
 Z_i^\text{f}
 & \longrightarrow \
 \begin{cases}
  \displaystyle
  \exp
  \left(
   - \frac{1}{\epsilon_1} \sum_{(x,\mu) \in \tilde{\cX}_i \times \cX_i^\text{f}} \Li_2 \left( q_2^{d_i} \frac{x}{\mu};p \right)
  \right)
  & (q_1 \to 1)
  \\[1.5em]
  \displaystyle
  \exp
  \left(
   - \frac{1}{d_i \epsilon_2} \sum_{(x,\mu) \in {\cX}_i \times \cX_i^\text{f}} \Li_2 \left( q_1 \frac{x}{\mu};p \right)
  \right)
  & (q_2 \to 1)  
 \end{cases}
\end{align}
\begin{align}
 Z_i^\text{af}
 & \longrightarrow \
 \begin{cases}
  \displaystyle
  \exp
  \left(
   \frac{1}{\epsilon_1} \sum_{(x,\mu) \in \tilde{\cX}_i \times \cX_i^\text{f}} \Li_2 \left( \frac{\mu}{x};p \right)
  \right)
  & (q_1 \to 1)
  \\[1.5em]
  \displaystyle
  \exp
  \left(
   \frac{1}{d_i \epsilon_2} \sum_{(x,\mu) \in {\cX}_i \times \cX_i^\text{f}} \Li_2 \left( \frac{\mu}{x};p \right)
  \right)
  & (q_2 \to 1)  
 \end{cases}
\end{align}
\end{subequations}
From these expressions, we can read off the effective twisted superpotential of 4d $\mathcal{N}=1$ theory on $\mathbb{R}^2 \times T^2$.
The saddle point equation of this 4d theory gives rise to the Bethe equation of the elliptic quantum integrable system.


\bibliographystyle{utphys}
\bibliography{NSL}

\end{document}